\documentclass[twocolumn]{aa}

\usepackage{graphicx}
\usepackage{amsmath,amsfonts,amssymb}
\usepackage{txfonts}
\usepackage{color}
\usepackage{natbib}
\usepackage{float}
\usepackage{dblfloatfix}
\usepackage{afterpage}
\usepackage{ifthen}
\usepackage[morefloats=12]{morefloats}
\usepackage{placeins}
\usepackage{multicol}
\bibpunct{(}{)}{;}{a}{}{,}
\usepackage[switch]{lineno}
\definecolor{linkcolor}{rgb}{0.6,0,0}
\definecolor{citecolor}{rgb}{0,0,0.75}
\definecolor{urlcolor}{rgb}{0.12,0.46,0.7}
\usepackage[breaklinks, colorlinks, urlcolor=urlcolor,
    linkcolor=linkcolor,citecolor=citecolor,pdfencoding=auto]{hyperref}
\hypersetup{linktocpage}

\def\setsymbol#1#2{\expandafter\def\csname #1\endcsname{#2}}
\def\getsymbol#1{\csname #1\endcsname}

\def\Planck{\textit{Planck}}

\newbox\tablebox    \newdimen\tablewidth
\def\leaderfil{\leaders\hbox to 5pt{\hss.\hss}\hfil}
\def\tablenote#1 #2\par{\begingroup \parindent=0.8em
    \abovedisplayshortskip=0pt\belowdisplayshortskip=0pt
    \noindent
    $$\hss\vbox{\hsize\tablewidth \hangindent=\parindent \hangafter=1 \noindent
    \hbox to \parindent{$^#1$\hss}\strut#2\strut\par}\hss$$
    \endgroup}

\def\L2{\ifmmode L_2\else $L_2$\fi}

\def\DeltaT{\ifmmode \Delta T\else $\Delta T$\fi}
\def\deltat{\ifmmode \Delta t\else $\Delta t$\fi}
\def\fknee{\ifmmode f_{\rm knee}\else $f_{\rm knee}$\fi}
\def\Fmax{\ifmmode F_{\rm max}\else $F_{\rm max}$\fi}
\def\solar{\ifmmode{\rm M}_{\mathord\odot}\else${\rm M}_{\mathord\odot}$\fi}
\def\Msolar{\ifmmode{\rm M}_{\mathord\odot}\else${\rm M}_{\mathord\odot}$\fi}
\def\Lsolar{\ifmmode{\rm L}_{\mathord\odot}\else${\rm L}_{\mathord\odot}$\fi}
\def\inv{\ifmmode^{-1}\else$^{-1}$\fi}
\def\mo{\ifmmode^{-1}\else$^{-1}$\fi}
\def\sup#1{\ifmmode ^{\rm #1}\else $^{\rm #1}$\fi}
\def\expo#1{\ifmmode \times 10^{#1}\else $\times 10^{#1}$\fi}
\def\,{\thinspace}
\def\lsim{\mathrel{\raise .4ex\hbox{\rlap{$<$}\lower 1.2ex\hbox{$\sim$}}}}
\def\gsim{\mathrel{\raise .4ex\hbox{\rlap{$>$}\lower 1.2ex\hbox{$\sim$}}}}

\def\simprop{\mathrel{\raise .4ex\hbox{\rlap{$\propto$}\lower 1.2ex\hbox{$\sim$}}}}
\def\deg{\ifmmode^\circ\else$^\circ$\fi}
\def\pdeg{\ifmmode $\setbox0=\hbox{$^{\circ}$}\rlap{\hskip.11\wd0 .}$^{\circ}
          \else \setbox0=\hbox{$^{\circ}$}\rlap{\hskip.11\wd0 .}$^{\circ}$\fi}
\def\arcs{\ifmmode {^{\scriptstyle\prime\prime}}
          \else $^{\scriptstyle\prime\prime}$\fi}
\def\arcm{\ifmmode {^{\scriptstyle\prime}}
          \else $^{\scriptstyle\prime}$\fi}
\newdimen\sa  \newdimen\sb
\def\parcs{\sa=.07em \sb=.03em
     \ifmmode \hbox{\rlap{.}}^{\scriptstyle\prime\kern -\sb\prime}\hbox{\kern -\sa}
     \else \rlap{.}$^{\scriptstyle\prime\kern -\sb\prime}$\kern -\sa\fi}
\def\parcm{\sa=.08em \sb=.03em
     \ifmmode \hbox{\rlap{.}\kern\sa}^{\scriptstyle\prime}\hbox{\kern-\sb}
     \else \rlap{.}\kern\sa$^{\scriptstyle\prime}$\kern-\sb\fi}
\def\ra[#1 #2 #3.#4]{#1\sup{h}#2\sup{m}#3\sup{s}\llap.#4}
\def\dec[#1 #2 #3.#4]{#1\deg#2\arcm#3\arcs\llap.#4}
\def\deco[#1 #2 #3]{#1\deg#2\arcm#3\arcs}
\def\rra[#1 #2]{#1\sup{h}#2\sup{m}}

\def\dots{\relax\ifmmode \ldots\else $\ldots$\fi}
\def\WHzsr{\ifmmode $W\,Hz\mo\,sr\mo$\else W\,Hz\mo\,sr\mo\fi}
\def\mHz{\ifmmode $\,mHz$\else \,mHz\fi}
\def\GHz{\ifmmode $\,GHz$\else \,GHz\fi}
\def\mKs{\ifmmode $\,mK\,s$^{1/2}\else \,mK\,s$^{1/2}$\fi}
\def\muKs{\ifmmode \,\mu$K\,s$^{1/2}\else \,$\mu$K\,s$^{1/2}$\fi}
\def\muKRJs{\ifmmode \,\mu$K$_{\rm RJ}$\,s$^{1/2}\else \,$\mu$K$_{\rm RJ}$\,s$^{1/2}$\fi}
\def\muKHz{\ifmmode \,\mu$K\,Hz$^{-1/2}\else \,$\mu$K\,Hz$^{-1/2}$\fi}
\def\MJysr{\ifmmode \,$MJy\,sr\mo$\else \,MJy\,sr\mo\fi}
\def\MJysrmK{\ifmmode \,$MJy\,sr\mo$\,mK$_{\rm CMB}\mo\else \,MJy\,sr\mo\,mK$_{\rm CMB}\mo$\fi}
\def\microns{\ifmmode \,\mu$m$\else \,$\mu$m\fi}

\def\muK{\ifmmode \,\mu$K$\else \,$\mu$\hbox{K}\fi}
\def\microK{\ifmmode \,\mu$K$\else \,$\mu$\hbox{K}\fi}
\def\muW{\ifmmode \,\mu$W$\else \,$\mu$\hbox{W}\fi}
\def\kms{\ifmmode $\,km\,s$^{-1}\else \,km\,s$^{-1}$\fi}
\def\kmsMpc{\ifmmode $\,\kms\,Mpc\mo$\else \,\kms\,Mpc\mo\fi}
\providecommand{\sorthelp}[1]{}

\def\WMAP{\textit{WMAP}}
\def\COBE{\textit{COBE}}
\def\LiteBIRD{\textit{LiteBIRD}}

\def\healpix{\texttt{HEALPix}}
\def\commander{\texttt{Commander}}

\def\commanderthree{\texttt{Commander3}}

\newcommand{\B}[0]{\tens{B}}

\renewcommand{\L}[0]{\tens{L}}

\newcommand{\M}[0]{\tens{M}}

\renewcommand{\P}[0]{\tens{P}}

\newcommand{\BP}{\textsc{BeyondPlanck}}

\newcommand{\mbeam}{\ensuremath{m_b}}

\newcommand{\cvar}{\ensuremath{c(\vartheta, \varphi, \psi)}}

\def\inv{^{-1}}

\begin{document}

\title{\bfseries{\scshape{BeyondPlanck}} VIII. Efficient Sidelobe Convolution and Corrections through Spin Harmonics}
%This author list corresponds to \title{Author list for L04\_CMB\_Foregrounds\_Extraction}
%Prepared by M. Lopez-Caniego (Marcos.Lopez.Caniego@sciops.esa.int), ESAC/ESA
%This version is from Thu Jul 12 18:11:48 2018 CET
%\subtitle{There are 152 co-authors in this list}
\newcommand{\nersc}[0]{1}
\newcommand{\princeton}[0]{2}
\newcommand{\helsinkiA}[0]{3}
\newcommand{\milanoA}[0]{4}
\newcommand{\triesteA}[0]{5}
\newcommand{\haverford}[0]{6}
\newcommand{\helsinkiB}[0]{7}
\newcommand{\triesteB}[0]{8}
\newcommand{\milanoB}[0]{9}
\newcommand{\milanoC}[0]{10}
\newcommand{\oslo}[0]{11}
\newcommand{\jpl}[0]{12}
\newcommand{\mpa}[0]{13}
\newcommand{\planetek}[0]{14}
\author{\small
M.~Galloway\inst{\oslo}\thanks{Corresponding author: M.~Galloway; \url{mathew.galloway@astro.uio.no}}
\and
M.~Reinecke\inst{\mpa}
\and
K.~J.~Andersen\inst{\oslo}
\and
\textcolor{black}{R.~Aurlien}\inst{\oslo}
\and
\textcolor{black}{R.~Banerji}\inst{\oslo}
\and
M.~Bersanelli\inst{\milanoA, \milanoB, \milanoC}
\and
S.~Bertocco\inst{\triesteB}
\and
M.~Brilenkov\inst{\oslo}
\and
M.~Carbone\inst{\planetek}
\and
L.~P.~L.~Colombo\inst{\milanoA}
\and
H.~K.~Eriksen\inst{\oslo}
\and
\textcolor{black}{M.~K.~Foss}\inst{\oslo}
\and
C.~Franceschet\inst{\milanoA,\milanoC}
\and
\textcolor{black}{U.~Fuskeland}\inst{\oslo}
\and
S.~Galeotta\inst{\triesteB}
\and
S.~Gerakakis\inst{\planetek}
\and
E.~Gjerl{\o}w\inst{\oslo}
\and
\textcolor{black}{B.~Hensley}\inst{\princeton}
\and
\textcolor{black}{D.~Herman}\inst{\oslo}
\and
M.~Iacobellis\inst{\planetek}
\and
M.~Ieronymaki\inst{\planetek}
\and
\textcolor{black}{H.~T.~Ihle}\inst{\oslo}
\and
J.~B.~Jewell\inst{\jpl}
\and
\textcolor{black}{A.~Karakci}\inst{\oslo}
\and
E.~Keih\"{a}nen\inst{\helsinkiA, \helsinkiB}
\and
R.~Keskitalo\inst{\nersc}
\and
G.~Maggio\inst{\triesteB}
\and
D.~Maino\inst{\milanoA, \milanoB, \milanoC}
\and
M.~Maris\inst{\triesteB}
\and
S.~Paradiso\inst{\milanoA, \milanoB}
\and
B.~Partridge\inst{\haverford}
\and
A.-S.~Suur-Uski\inst{\helsinkiA, \helsinkiB}
\and
T.~L.~Svalheim\inst{\oslo}
\and
D.~Tavagnacco\inst{\triesteB, \triesteA}
\and
H.~Thommesen\inst{\oslo}
\and
D.~J.~Watts\inst{\oslo}
\and
I.~K.~Wehus\inst{\oslo}
\and
A.~Zacchei\inst{\triesteB}
}
\institute{\small
Computational Cosmology Center, Lawrence Berkeley National Laboratory, Berkeley, California, U.S.A.\goodbreak
\and
Department of Astrophysical Sciences, Princeton University, Princeton, NJ 08544,
U.S.A.\goodbreak
\and
Department of Physics, Gustaf H\"{a}llstr\"{o}min katu 2, University of Helsinki, Helsinki, Finland\goodbreak
\and
Dipartimento di Fisica, Universit\`{a} degli Studi di Milano, Via Celoria, 16, Milano, Italy\goodbreak
\and
Dipartimento di Fisica, Universit\`{a} degli Studi di Trieste, via A. Valerio 2, Trieste, Italy\goodbreak
\and
Haverford College Astronomy Department, 370 Lancaster Avenue,
Haverford, Pennsylvania, U.S.A.\goodbreak
\and
Helsinki Institute of Physics, Gustaf H\"{a}llstr\"{o}min katu 2, University of Helsinki, Helsinki, Finland\goodbreak
\and
INAF - Osservatorio Astronomico di Trieste, Via G.B. Tiepolo 11, Trieste, Italy\goodbreak
\and
INAF-IASF Milano, Via E. Bassini 15, Milano, Italy\goodbreak
\and
INFN, Sezione di Milano, Via Celoria 16, Milano, Italy\goodbreak
\and
Institute of Theoretical Astrophysics, University of Oslo, Blindern, Oslo, Norway\goodbreak
\and
Jet Propulsion Laboratory, California Institute of Technology, 4800 Oak Grove Drive, Pasadena, California, U.S.A.\goodbreak
\and
Max-Planck-Institut f\"{u}r Astrophysik, Karl-Schwarzschild-Str. 1, 85741 Garching, Germany\goodbreak
\and
Planetek Hellas, Leoforos Kifisias 44, Marousi 151 25, Greece\goodbreak
}

\authorrunning{Galloway et. al.}
\titlerunning{Sidelobe Corrections}

\abstract{We introduce a new formulation of the \texttt{Conviqt} convolution algorithm in terms of spin harmonics, and apply this to the problem of sidelobe correction for \BP, the first end-to-end Bayesian Gibbs sampling framework for CMB analysis. We compare our implementation to the previous \Planck\ LevelS implementation, and find good agreement between the two codes in terms of accuracy, but with a speed-up reaching a factor of 3--10, depending on the frequency bandlimits, $l_{\textrm{max}}$ and $m_{\textrm{max}}$. The new algorithm is significantly simpler to implement and maintain, since all low-level calculations are handled through an external spherical harmonic transform library. We find that our mean sidelobe estimates for \Planck\ LFI agree well with previous efforts. Additionally, we present novel sidelobe rms maps that quantify the uncertainty in the sidelobe corrections due to variations in the sky model.}


\maketitle

\section{Introduction}
\label{sec:introduction}

One of the important systematic effects that must be accounted for in CMB instruments is telescope stray light and sidelobes \citep[e.g.,][]{barnes2003,planck2013-p02a}. This is the non-zero response of the detector to areas of the sky outside the main beam, however that is defined. Because microwave telescopes typically work near their diffraction limit, some level of sidelobes is inevitable. Furthermore, their structure can be complicated by many different physical effects, such as spurious optical reflections or manufacturing irregularities in the detectors or optical elements. These signal contributions can have far-reaching consequences on the observed signal, particularly at large angular scales, as they do not behave in the same sky-stationary manner as the main beam signal.

Sidelobe signals can produce many types of errors in CMB analysis pipelines, and they represent a potent source of systematic contamination \citep[e.g.,][]{planck2014-a04,bp17}. In particular, as the sidelobe response functions often are broadly distributed, this contamination can confuse important signals like the CMB solar and orbital dipoles that are used for calibration. Sidelobes uncertainties couple directly with foreground emission from diffuse galactic components, producing an important source of contamination. In some experiments a further contaminating signal can originate from a source not on the sky, such as ground pickup or radio-frequency (RF) noise. In all cases, sidelobe signal is detrimental to the quality of the final sky maps and parameter estimates, and requires a dedicated effort to remove it. Characterizing and correcting these spurious signals is therefore an important part of optimal CMB mapmaking, and requires optimized algorithms to characterize them efficiently. One of the most important of these is to convolve a beam or sidelobe response function with a sky map or model to generate a re-observed map. 

Full sky convolution on the sphere is a problem that has been important in the CMB field since the earliest satellite measurements. Early experiments like \COBE\ \citep{cobe_sl} either did not model sidelobes at all, or used simple pixel-based convolution approaches which even for their low resolution required radially symmetric beam approximations \citep{radialapprox}, or limited the applications to large scales \citep{burigana2001}.

\citet{Wandelt:2001} presented the first harmonic space convolution algorithm, often referred to as ``total convolution'', which achieved a large performance gain over pixel-based methods, by as much as a factor of $O(\sqrt{N_\mathrm{pix}})$. This breakthrough allowed the calculation of these convolutions easily enough that they could be applied to each simulation, instead of requiring a dedicated study necessitating months of runtime.

Next, \citet{conviqt} developed the \texttt{Conviqt} approach, which was used both by several official \Planck\ analysis pipelines (\citealt{planck2014-a04}, \citealt{planck2014-a10}, \citealt{npipe}) and to generate the \Planck\ Full Focal Plane (FFP) simulations \citep{planck2014-a14}. This approach was an improvement over the state of the art, speeding up the computation of the Wigner recursion relationships used in the original harmonic space algorithm, as well as providing a standardized, user friendly library, \texttt{libconviqt}, that was incorporated into numerous pipelines. 

In this paper, we introduce a new formulation of the \texttt{Conviqt} algorithm that is based on Spherical Harmonic Transforms (SHTs), rather than directly computing the Wigner matrix elements. We are thus able to leverage the highly optimized \texttt{libsharp} SHT library to perform the bulk of the calculations \citep{libsharp}. Although this new approach was not developed specifically for \BP, this paper is the first to explicitly derive, discuss, and benchmark the method.

\section{Sidelobes, \texttt{libconviqt} and \texttt{libsharp}}

\subsection{Total Convolution through Spin Harmonics}

\label{sec:conviqt}

Given a sky map, $s(\hat{n})$, and beam, $b(\hat{n})$, our task is to
compute a quantity $\cvar \in \mathbb{R}$ that represents the
convolution of these two fields, with the beam oriented in polar
coordinates $(\vartheta, \varphi)$,\footnote{In this paper, $\vartheta$ and $\varphi$ are the co-latitude and longitude of a location on the sphere, i.e., they have the same meaning as in the \healpix\ context \citep{gorski2005}.} and rotated around its own central
axis by $\psi$,
\begin{equation}
\begin{aligned}
  \cvar &:= s_{lm_s} * b_{lm_b}\\ 
   &\equiv \int_{4\pi} s(\hat{n})
  b\big(\hat{n}'(\vartheta,\varphi)-\hat{n},\psi\big)\, d\Omega_{\hat{n}}.
  \label{eq:convolution}
\end{aligned}
\end{equation}
Here, $s_{lm_s}$ denotes the spherical harmonic coefficients of the sky signal, and $b_{lm_b}$ is the beam in the same representation. In this expression, care must been taken to distinguish between the sky and beam bandlimits, $m_s$ and $m_b$, as the two indices will be treated separately in the following derivation.

A computationally efficient solution for this problem was derived by
\citet{conviqt}, who exploited fast recurrence relations for Wigner
$d$ matrix elements to evaluate Eq.~\eqref{eq:convolution} in harmonic
space. In the following, we will show that this equation
can alternatively be expressed in terms of spin-harmonics. The
resulting algebra is in principle identical to the recursion relations
used by \citet{conviqt}, but the equations are simply repackaged
in a format that is significantly easier to implement in practical
computer code, since it may use existing and highly optimized
spherical harmonic libraries, such as \texttt{libsharp} \citep{libsharp}, to perform
the computationally expensive parts.

As shown by \citet{Wandelt:2001}, Eq.~(\ref{eq:convolution}) can be
evaluated efficiently in harmonic space as
\begin{equation}
\cvar = \sum_{l,m_s,m_b} s_{lm_s} b^\ast_{lm_b}
[D^{l}_{m_sm_b}(\varphi,\vartheta,\psi)]^\ast,
\label{eq:totalconvolver}
\end{equation}
where $s_{lm_s}$ and $b_{lm_b}$ are the spherical harmonic
coefficients of the signal and beam, respectively, and
$D^{l}_{m_sm_b}$ is the Wigner $D$-matrix.

This may be expressed as \citep{goldberg}
\begin{equation}
D^l_{-ms}(\varphi,\vartheta,-\psi) = (-1)^m\sqrt{\frac{4\pi}{2l+1}}
{}_sY_{lm}(\vartheta,\varphi) e^{is\psi},
\end{equation}
where ${}_sY_{lm}(\vartheta,\varphi)$ is the spin-weighted
spherical harmonic and the placement of the negative signs are an arbitrary historical convention. Inserting this expression into
Eq.~(\ref{eq:totalconvolver}) yields
\begin{equation}
\cvar = \sum_{l,m_s,m_b}\sqrt{\frac{4\pi}{2l+1}} s_{lm_s} b_{l-m_b}\, \cdot {}_{-m_b}Y_{lm_s}(\vartheta,\varphi) e^{im_b\psi},
\end{equation}
where we have assumed that the beam is real-valued in position space,
and we have used the symmetry relations,
\begin{align}
  D^l_{-m_s,-m_b}(\vartheta,\varphi)&=(-1)^{m_s+m_b}[D^{l}_{m_s,m_b}(\vartheta,\varphi)]^\ast,\\
  b^\ast_{l,m_b} (-1)^{m_b}&=b_{l,-m_b}.\label{eq:breal}
\end{align}
Pulling the summation over $m_b$ in front of the other sums yields
\begin{equation}
\cvar = \sum_{m_b} e^{im_b\psi} \sum_{l,m_s} \sqrt{\frac{4\pi}{2l+1}}
s_{lm_s} b_{l-m_b}\, \cdot {}_{-m_b}Y_{lm_s}(\vartheta,\varphi).
\label{eq:c_m}
\end{equation}
The terms in this outer sum can be arranged in the form $m_b=0, \pm 1, \pm 2, \dots$.
The contribution from \mbox{$m_b=0$} can be interpreted as a spin-0
spherical harmonic transform of the quantity $\sqrt{4\pi/(2l+1)}
s_{lm_s} b_{l0}$, which can be easily computed by a library like
\texttt{libsharp} \citep{libsharp}.

Since $\cvar \in \mathbb{R}$, we know that the contributions from the pairs
$m_b=\pm 1, \pm 2, \dots$ must be complex conjugate with respect to each other, and their combined
contribution is therefore 
\begin{align}
e^{im_b\psi}{}_{m_b}&S(\vartheta,\varphi) + e^{-im_b\psi}{}_{m_b}S^\ast(\vartheta,\varphi) =
 \\\nonumber
 &2\left[\cos(m_b\psi)\text{Re}({}_{m_b}S(\vartheta,\varphi)) -
 \sin(m_b\psi)\text{Im}({}_{m_b}S(\vartheta,\varphi))\right],
\end{align}
where we have defined
\begin{equation}
{}_{m_b}S(\vartheta,\varphi) \equiv \sum_{l,m_s} \sqrt{\frac{4\pi}{2l+1}} s_{lm_s}
b_{l-m_b}\, \cdot {}_{-m_b}Y_{lm_s}(\vartheta,\varphi).
\label{eq:S}
\end{equation}
This is a spherical harmonic transform of a quantity with spin $m_b$,
which can also be computed efficiently by \texttt{libsharp}.

In practice, the transforms in Eq.~(\ref{eq:S}) are implemented by
separating $S$ into its gradient and curl (or $E$ and $B$)
coefficients, $a_{lm}$ \citep{lewis_2005}, 
\begin{equation}
{}_{m_b}S_{lm_s} = -\left({}_{m_b}E_{lm_s} + i\,{}_{m_b}B_{lm_s}\right),
\end{equation}
using the symmetry relations
${}_{m_b}E_{l-m_s}=(-1)^{m_s}{}_{m_b}E_{lm_s}^*$ and
${}_{m_b}B_{l-m_s}=(-1)^{m_s}{}_{m_b}B_{lm_s}^*$, and the overall minus sign is a convention. Again making use of the symmetry relation
in Eq.~\eqref{eq:breal}, this results in
\begin{align}
  {}_{m_b}E_{l,m_s} &= -s_{lm_s} \text{Re}(b_{l,\mbeam}) \label{spin1}\\
  {}_{m_b}B_{l,m_s} &= -s_{lm_s} \text{Im}(b_{l,\mbeam}) \label{spin2}.
\end{align}

To summarize, efficient evaluation of the convolution integral in
Eq.~(\ref{eq:convolution}) may be done through the following steps:
\begin{enumerate}
  \item For each $m = {0,\ldots, m_{b}}$, pre-compute the spin spherical
    harmonic coefficients in Eqs.~\eqref{spin1}--\eqref{spin2}, and
    compute the corresponding spin-$m_{b}$ SHT
    with an external library such as \texttt{libsharp}; this results
    in a three-dimensional data cube of the form
    $c(\vartheta,\varphi,m_{b})$.
  \item For each position on the sky, $(\vartheta,\varphi)$, perform a
    Fourier transform to convert these coefficients to
    $c(\vartheta,\varphi,\psi)$, as given by Eq.~\eqref{eq:c_m}.
\end{enumerate}
In practice, the resulting $c(\vartheta,\varphi,\psi)$ data object is
evaluated at a finite pixel resolution typically set to match the beam
bandlimit. To obtain smooth estimates within this data object, a wide
range of interpolation schemes may be employed, trading off
computational efficiency against accuracy. This issue is
identical to previous approaches \citep{Wandelt:2001,conviqt},
and we refer the interested reader to those papers for further details.

\subsection{Comparison with \texttt{libconviqt}}

\begin{figure}[t]
  \center
  \includegraphics[width=\linewidth]{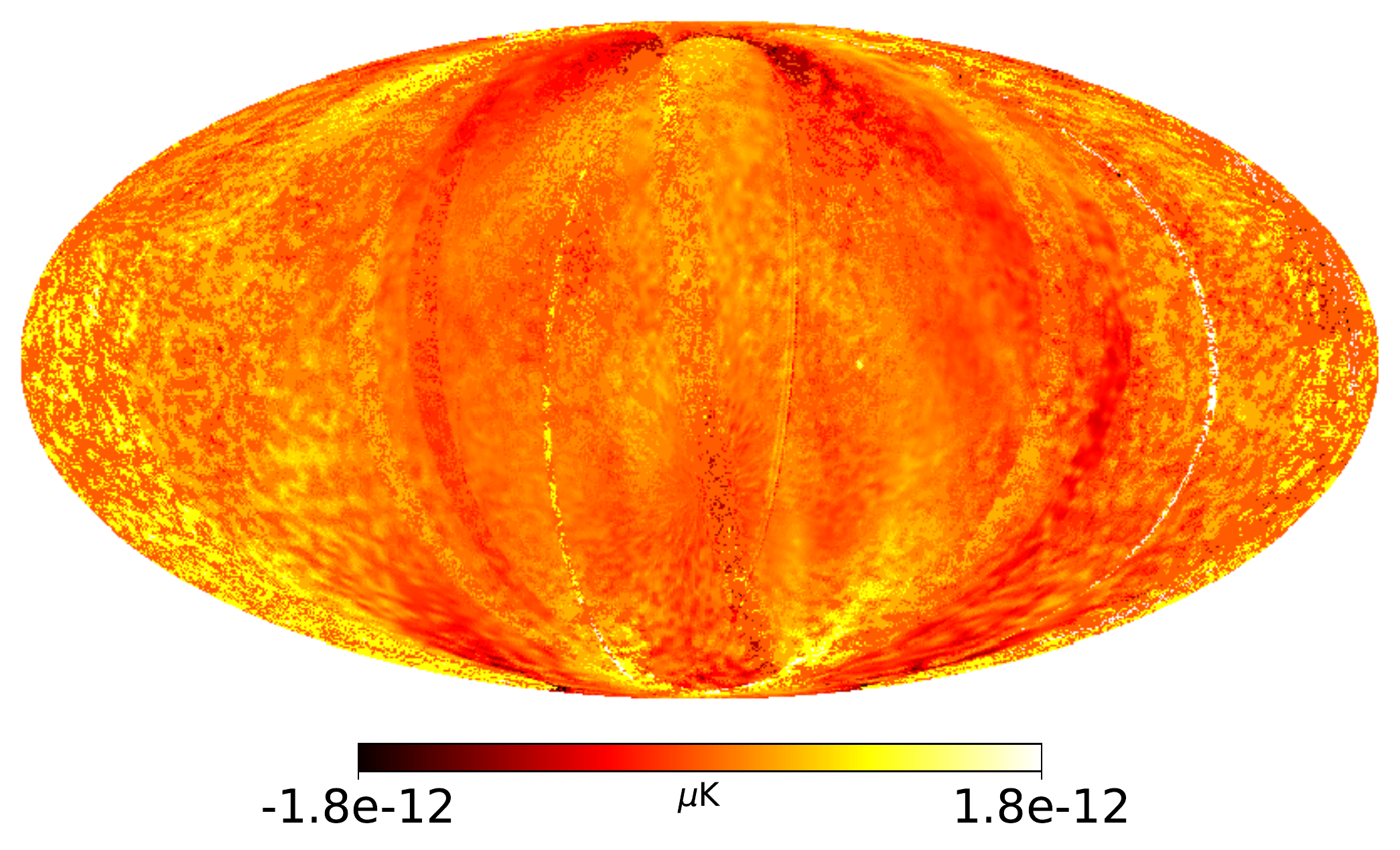}
  \caption{Map level difference in temperature of the new SHT convolution algorithm compared to the old \texttt{Conviqt} approach, observed using the identical pointing of the first year of the 30\,GHz \Planck\ detectors. The differences are at the level of machine precision, indicating full agreement between the two algorithms. 
  }\label{fig:differences}
\end{figure}

To compare the results of this new total convolution approach with the older \texttt{libconviqt} approach of \citet{conviqt}, we evaluate the convolution between the beam for one of the LFI 30\,GHz receivers (28M) and a \commander\ 30\,GHz sky model \citep{bp13,bp14} using both methods. The resulting convolution cubes are then observed using LFI's scanning strategy for the first year of the \Planck\ flight. The resulting map differences are shown in Fig.~\ref{fig:differences}. The convolution cubes are also directly compared for accuracy, and found to agree with an integrated difference at the $10^{-8}$ level, indicative of differences at the level of numerical precision.

\begin{figure}[t]
  \center
  \includegraphics[width=\linewidth]{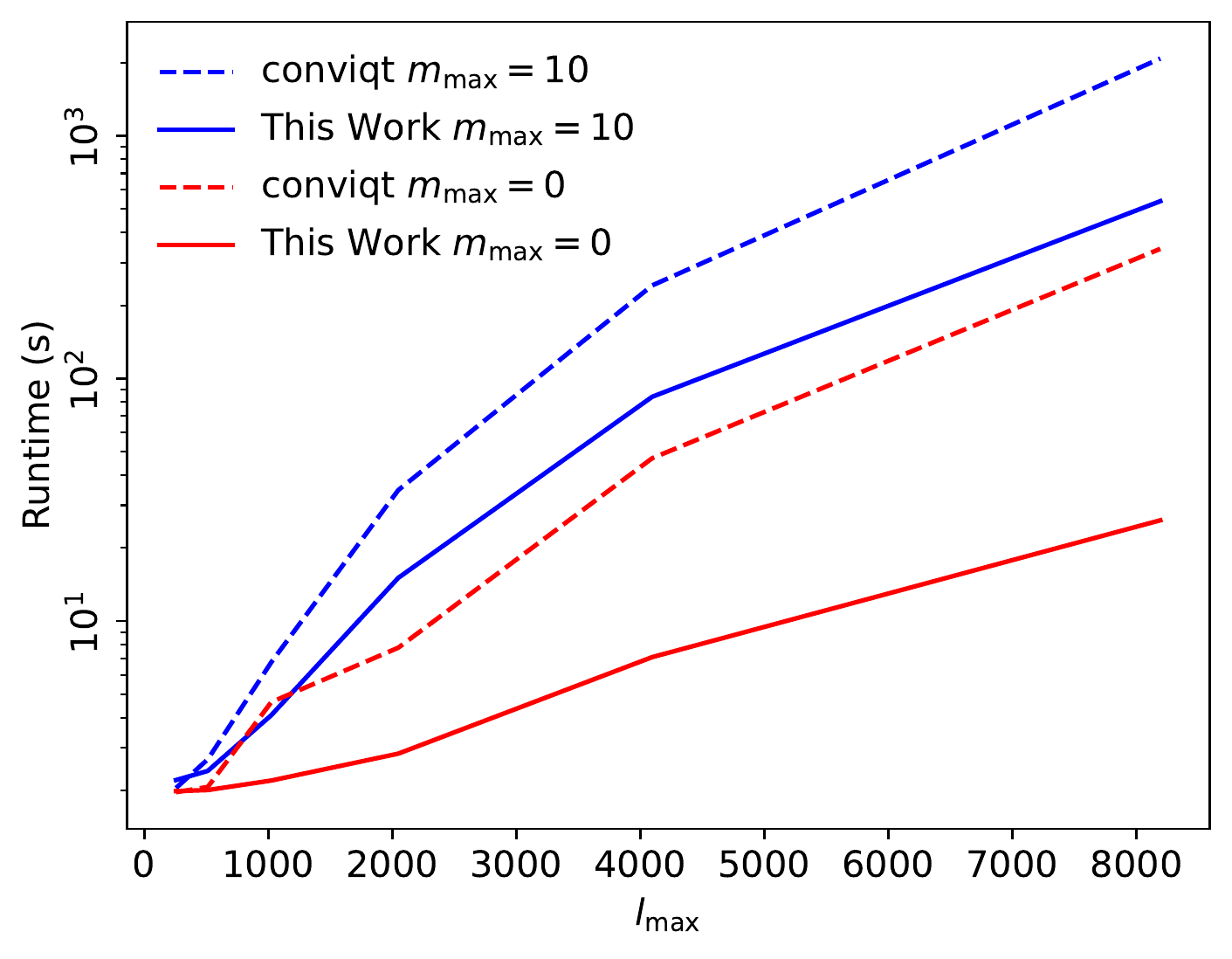}
	\caption{Runtime comparison between the \texttt{libconviqt} approach and the new spin-SHT approach for the convolution of an elliptical Gaussian with a set of random sky $a_{l,m}$s. This work ties or outperforms the previous approach for all values of $l_\mathrm{max}$ from 256 to 8192 for both $m_\mathrm{max}$ values shown. Note the log scale on the y-axis.
  }\label{fig:speed}
\end{figure}

Figure~\ref{fig:speed} compares the runtime between the two approaches for a test configuration with an elliptical beam and a fixed sky model, and with $m_\mathrm{max}=0$ (using only the radially symmetric part of the beam) and $m_\mathrm{max}=10$, respectively. In both cases, the new implementation outperforms the old approach at all but the lowest $l_\mathrm{max}$, where the data read time dominates. Additionally, for compatibility with the old \texttt{libconviqt} approach, this test was performed with an older version of \texttt{libsharp}, so we expect that the new algorithm scales even more favourably than this with the latest implementation. We note that this is a significant real-life advantage of the new approach: any improvement in SHT libraries, which typically are subject to intensive algorithm development and code maintenance, translates directly into a computational improvement for the convolution algorithm.

\section{Sidelobe Models}

\begin{figure*}[t]
  \center
  \includegraphics[width=0.33\linewidth]{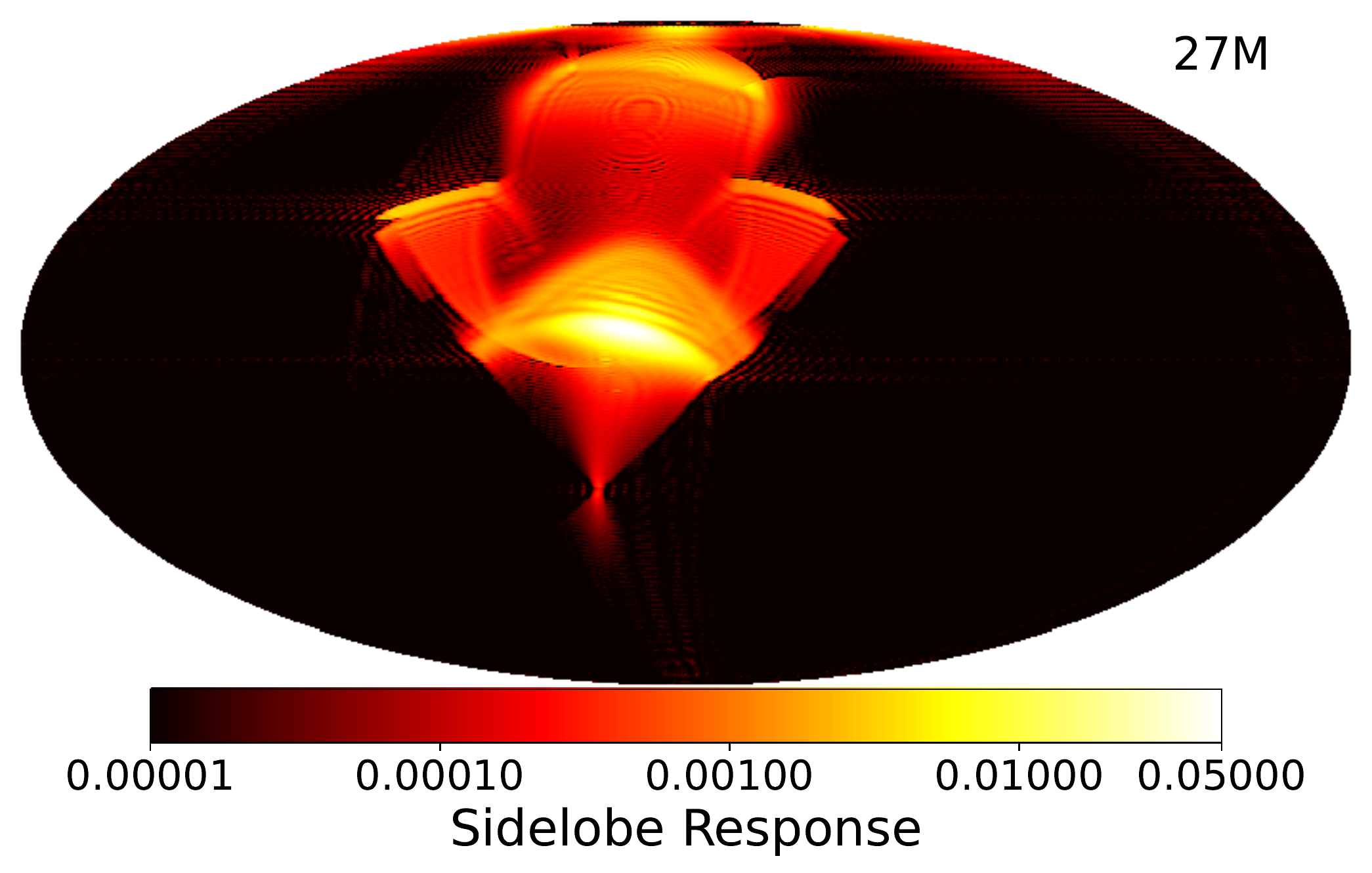}
  \includegraphics[width=0.33\linewidth]{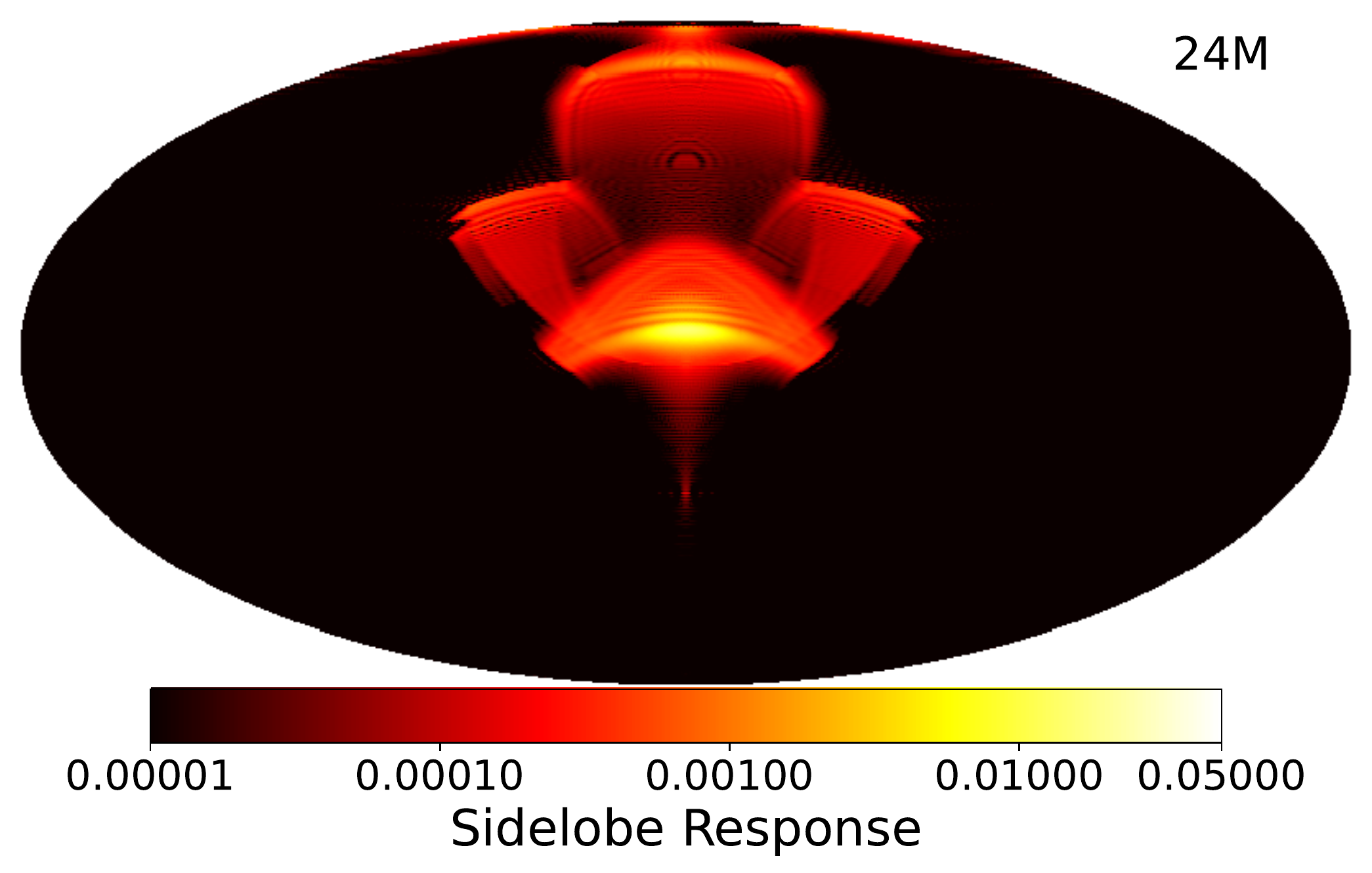}
  \includegraphics[width=0.33\linewidth]{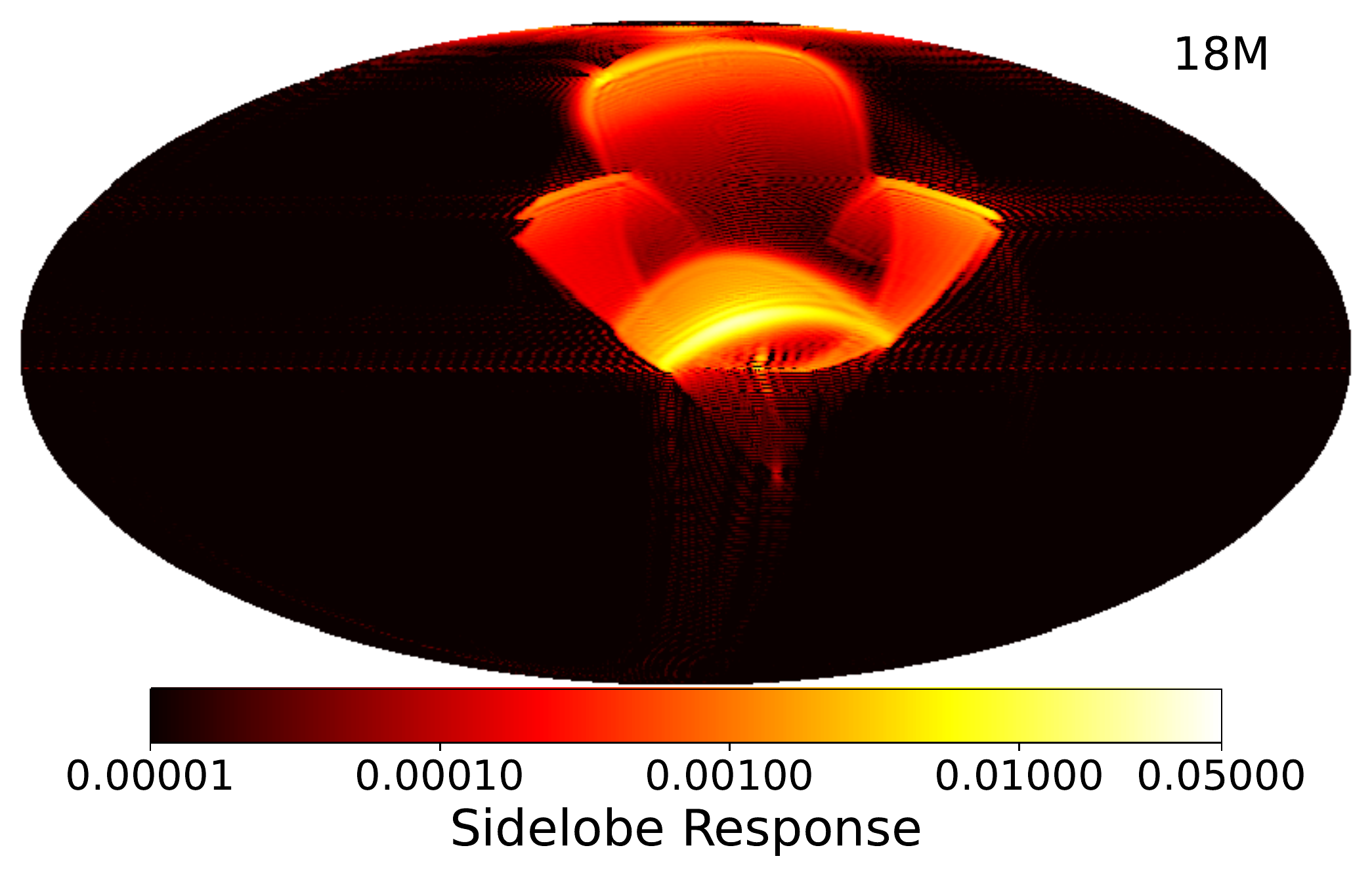}\\
  \caption{Maps of the sidelobe response on the sky from a representative detector at (left to right) 30\,GHz, 44\,GHz and 70\,GHz. The beam orientation is such that the main beam is pointed directly at the north pole in these maps. The intensities are normalized such that the main beams have unit power at $l=0$.
  }\label{fig:slresponse}
\end{figure*}

Figure~\ref{fig:slresponse} shows characteristic sidelobe response functions evaluated at a fixed frequency on the sky for a detector in each Planck LFI band. The sidelobe response for each detector within a single \Planck\ band look visually quite similar, so only these representative ones are shown here. Each is stored on disk as a set of $a_{lm}$'s with $l_\mathrm{max} = 512$ and $m_\mathrm{max} = 100$. 

\subsection{Main Beam Treatment}

In the \BP\ analysis, the sidelobe and main beam components of the sky response are separated, and the sidelobes are treated as a nuisance signal similar to the orbital dipole correction term, as can be seen in the \BP\ global parametric model of the data:
\begin{equation}
\begin{split}
d_{j,t} = g_{j,t}&\left[\P_{tp,j}\B_{pp',j}\sum_{c}
\M_{cj}(\beta_{p'}, \Delta bp^{j})a^c_{p'}  + s^{\mathrm{orb}}_{j,t}  
+ s^{\mathrm{fsl}}_{j,t}\right] + \\
+ &s^{\mathrm{1hz}}_{j,t} + n^{\mathrm{corr}}_{j,t} + n^{\mathrm{w}}_{j,t}.
\end{split}
\label{eq:datamodel}
\end{equation}
The other terms in this equation are discussed in detail in \cite{BP01}, but here the main beam signal is denoted as $B_{pp',j}$ and the sidelobe signal is extracted from the signal contribution and expressed as $s^{\mathrm{fsl}}_{j,t}$. This distinction allows the sidelobes to be treated separately from the main beam in all respects. Treating the main beam using the \texttt{Conviqt} formalism of this paper would be possible, but the additional precision needed to model it accurately would require much higher $l_\mathrm{max}$, and therefore greatly increased computational time and memory requirements. 

In the \BP\ analysis, the main beam is used (in conjunction with the sidelobes) to compute the full 4$\pi$ dipole response, as detailed in Sect.~\ref{sec:dipole}. Additionally, a Gaussian main beam approximation is used during component separation to smooth the sky model to the appropriate beam resolution for each channel. During mapmaking, beam effects are ignored and the beam is assumed to be pointed at the center of each pixel. 

\subsection{Sidelobe Normalization}
\label{sec:normalization}

We adopt a normalization of the sidelobes that differs slightly from the normalization used within the \Planck\ LFI collaboration. The \Planck\ 2018 LFI beam products leave a small portion (around 1\%) of known missing power within the system unassigned due to uncertainties about to which component it should be assigned \citep{planck2014-a05}. In the current analysis, we rather adopt the same approximation as for \Planck\ DR4 \citep{npipe}, and renormalize the beam transfer function such that this power is distributed proportionally at each $l$; that is, we rescale the beam transfer function $B_l$ such that its full sky integral is $B_0 = 1$. This re-scaling is equivalent to assigning the unknown beam power uniformly over the entire beam. We note that this normalization is in either case always done before any higher-level analysis for both \Planck\ 2018 and DR4; the only difference is whether the renormalization must be performed by external users through deconvolution of a non-unity normalized main beam transfer function or not. 

\subsection{Orbital Dipole and Quadrupole Sidelobe Response}
\label{sec:dipole}

\begin{figure*}[t]
  \center
  \includegraphics[width=0.33\linewidth]{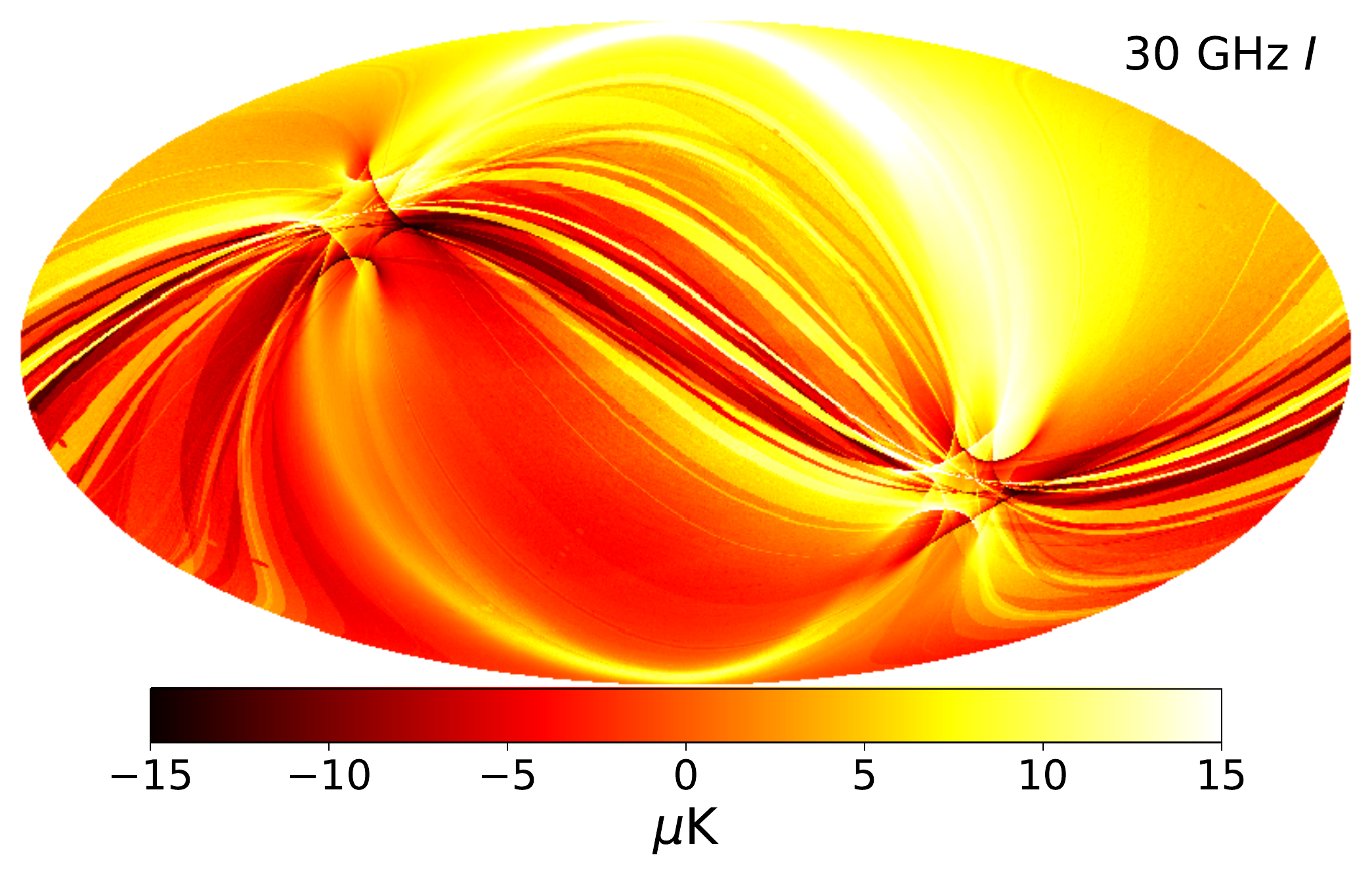}
  \includegraphics[width=0.33\linewidth]{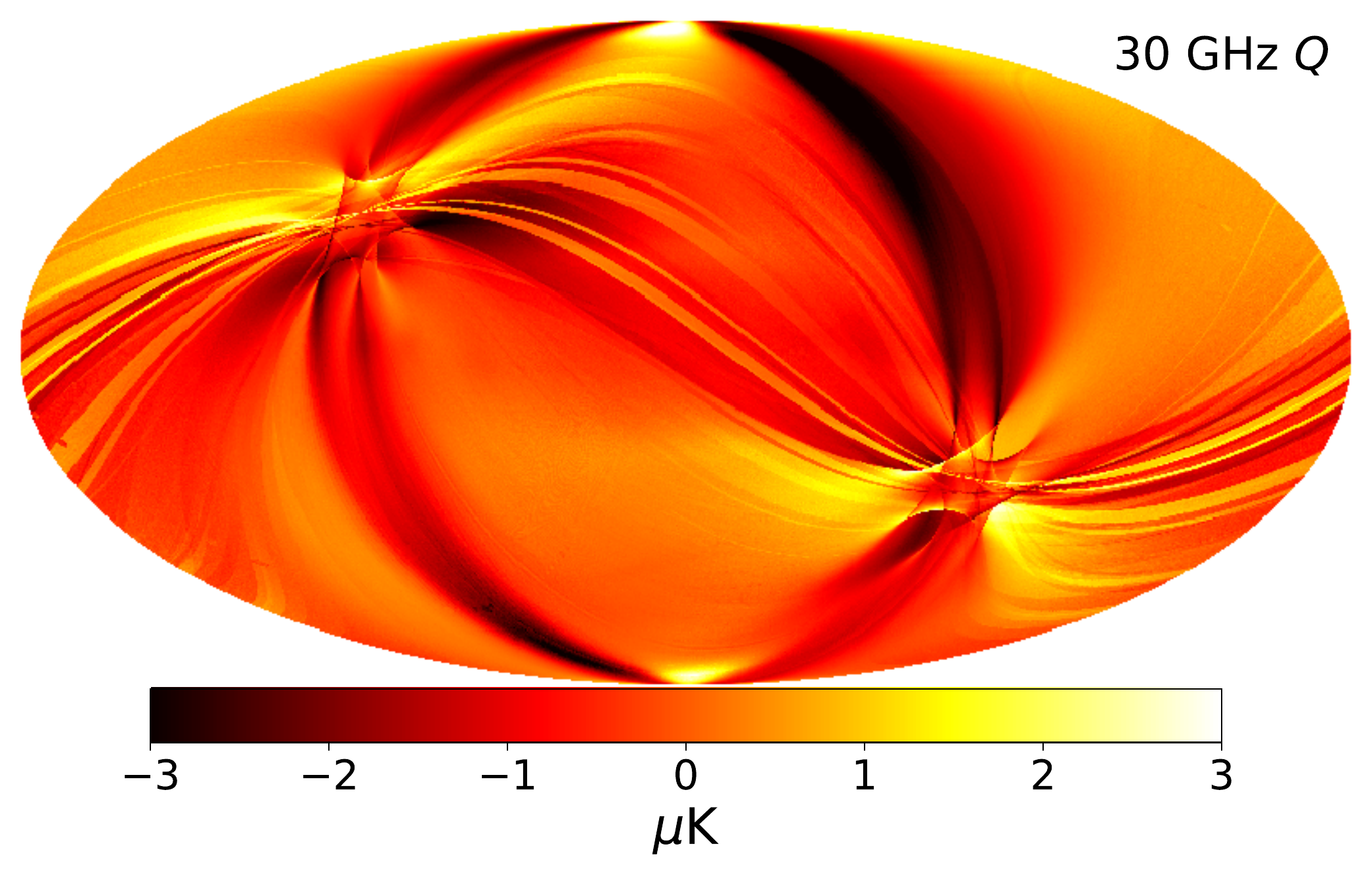}
  \includegraphics[width=0.33\linewidth]{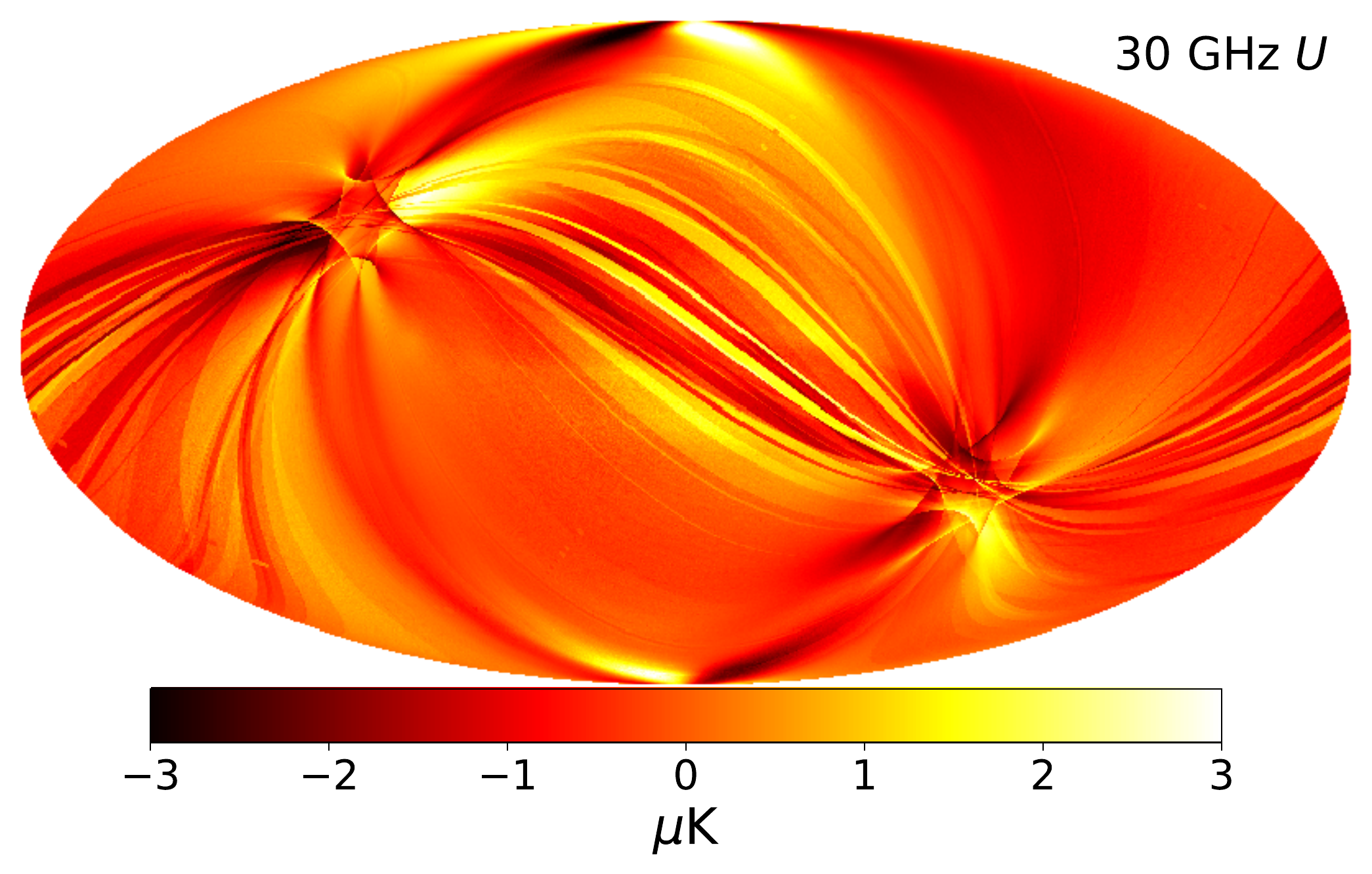}\\
    \includegraphics[width=0.33\linewidth]{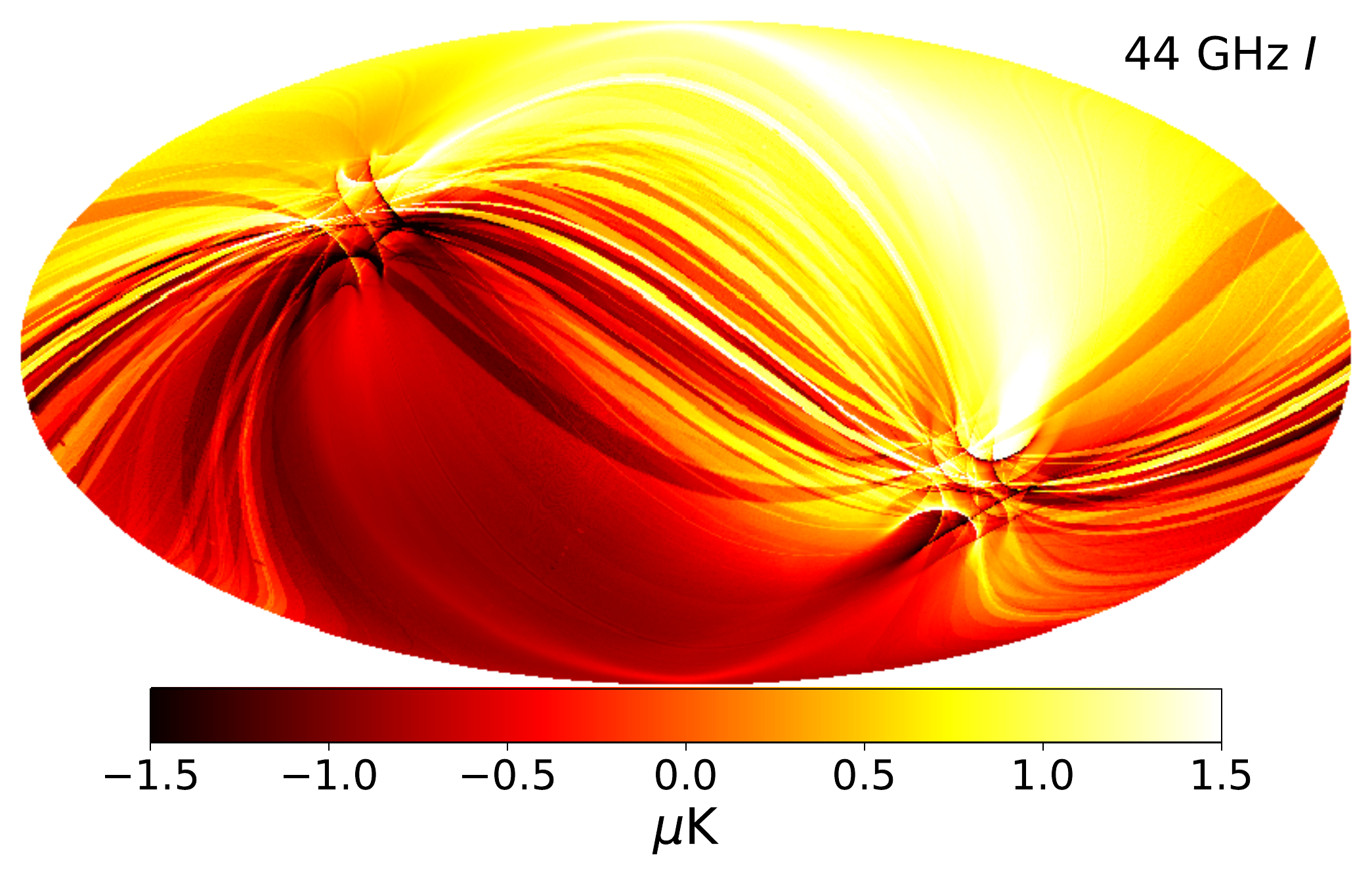}
  \includegraphics[width=0.33\linewidth]{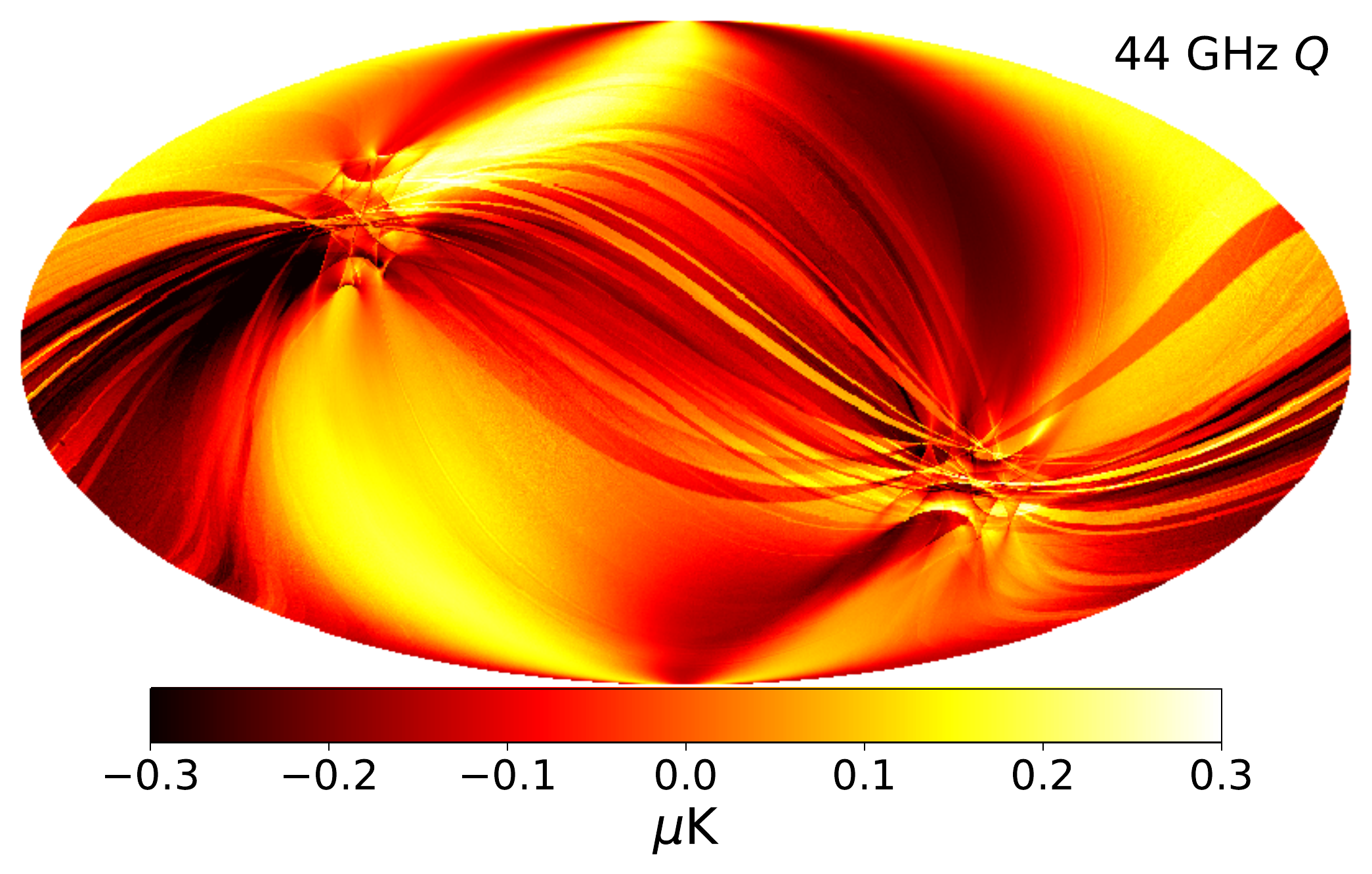}
  \includegraphics[width=0.33\linewidth]{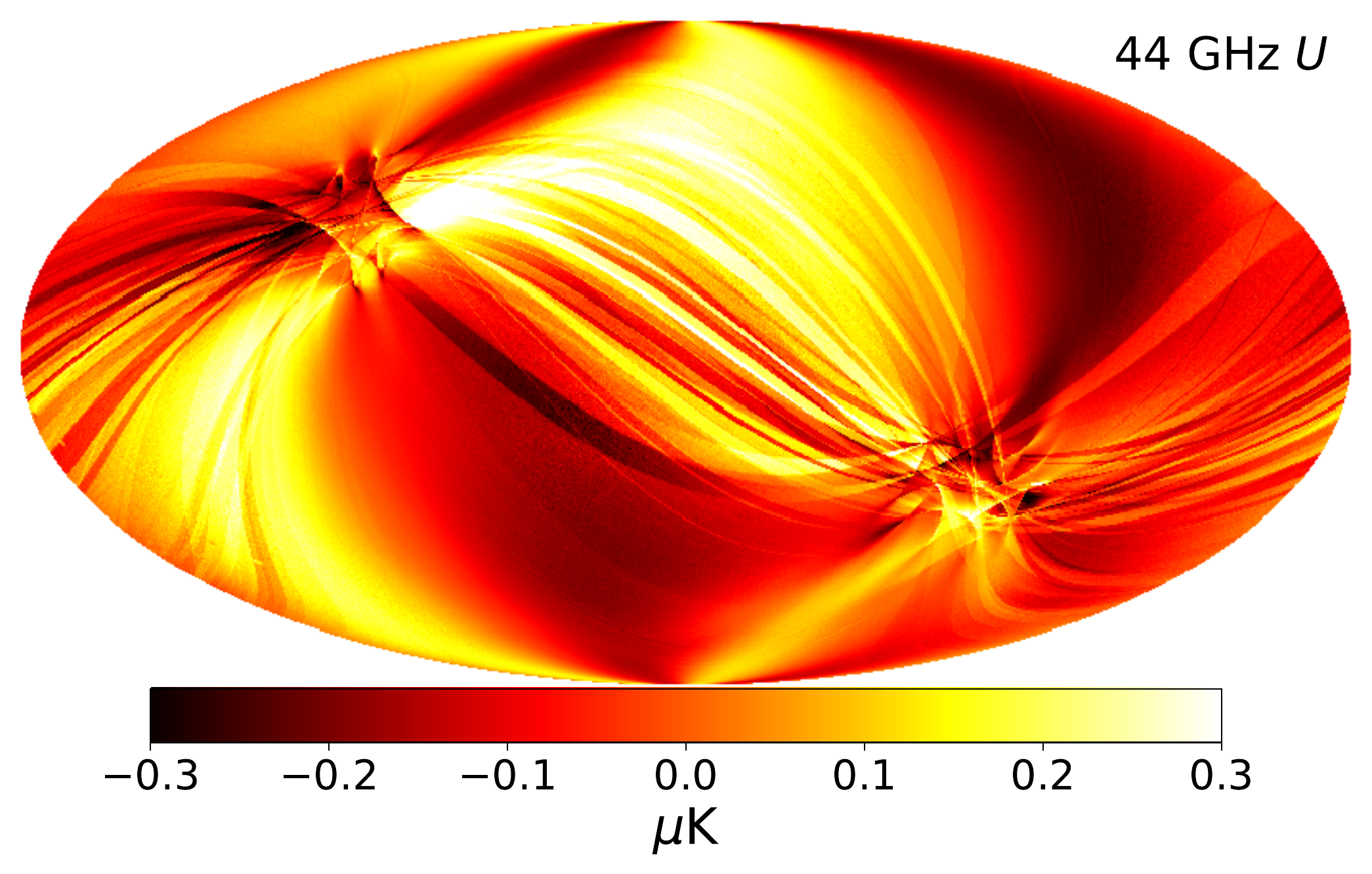}\\
    \includegraphics[width=0.33\linewidth]{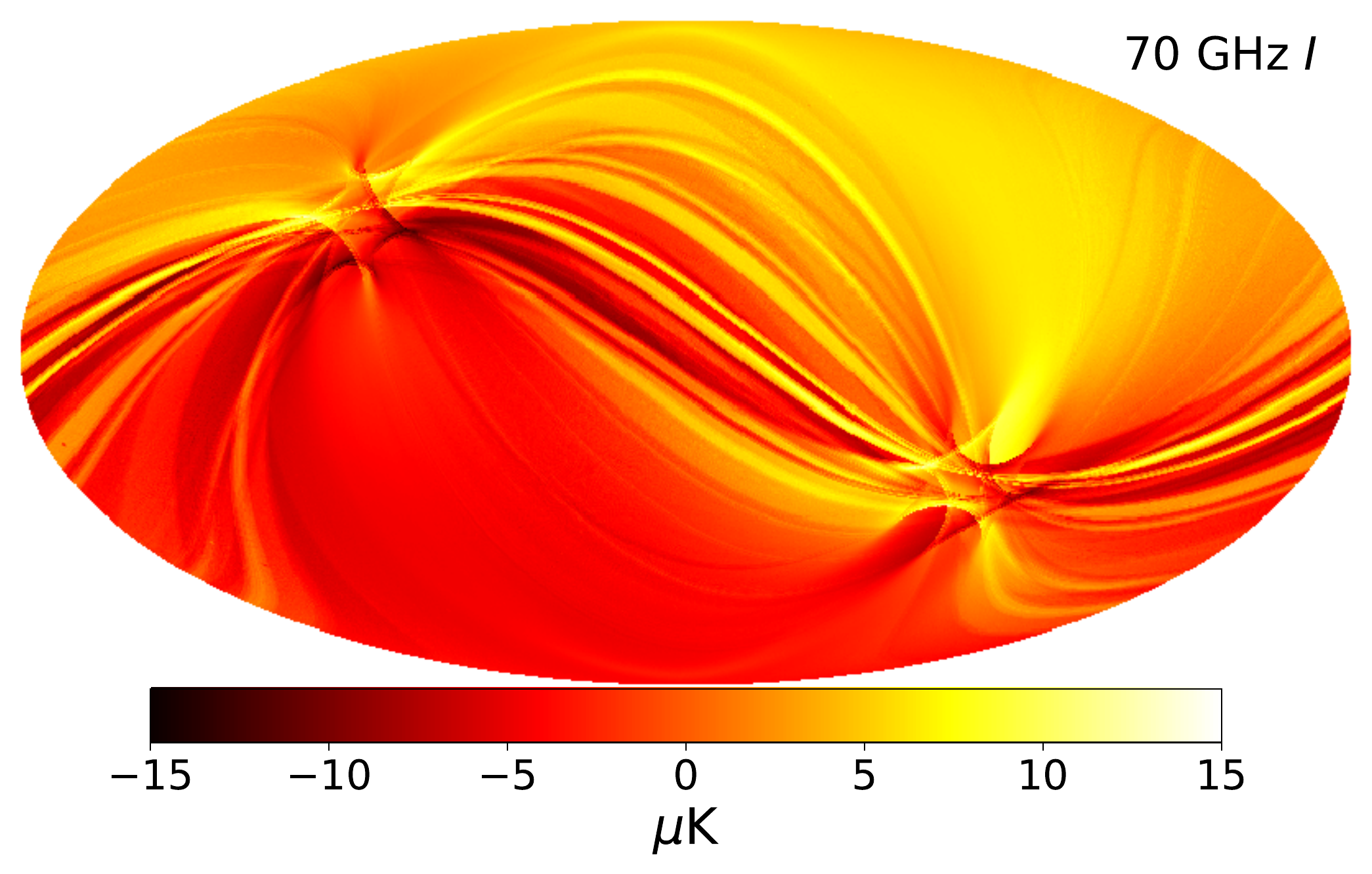}
  \includegraphics[width=0.33\linewidth]{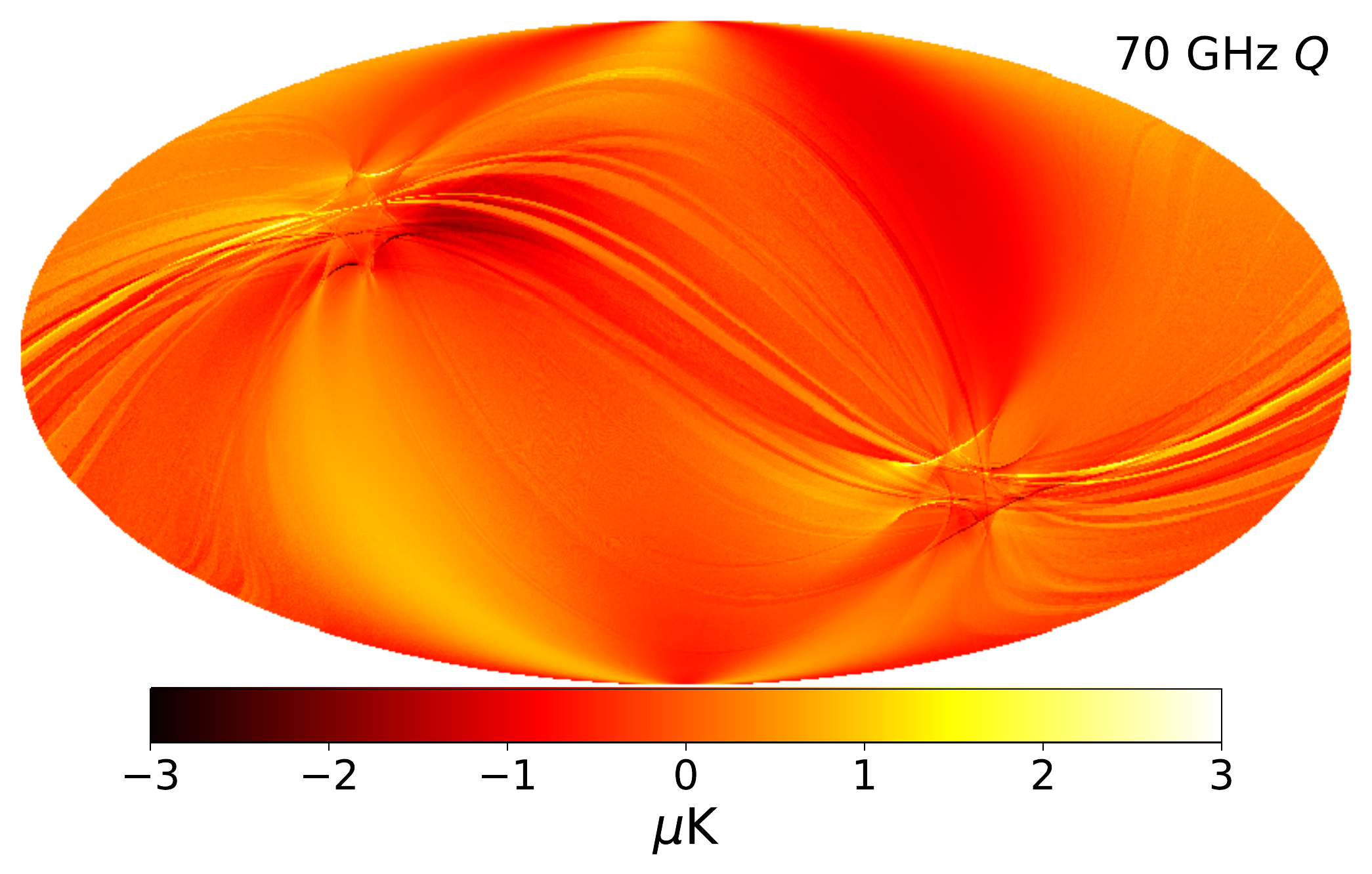}
  \includegraphics[width=0.33\linewidth]{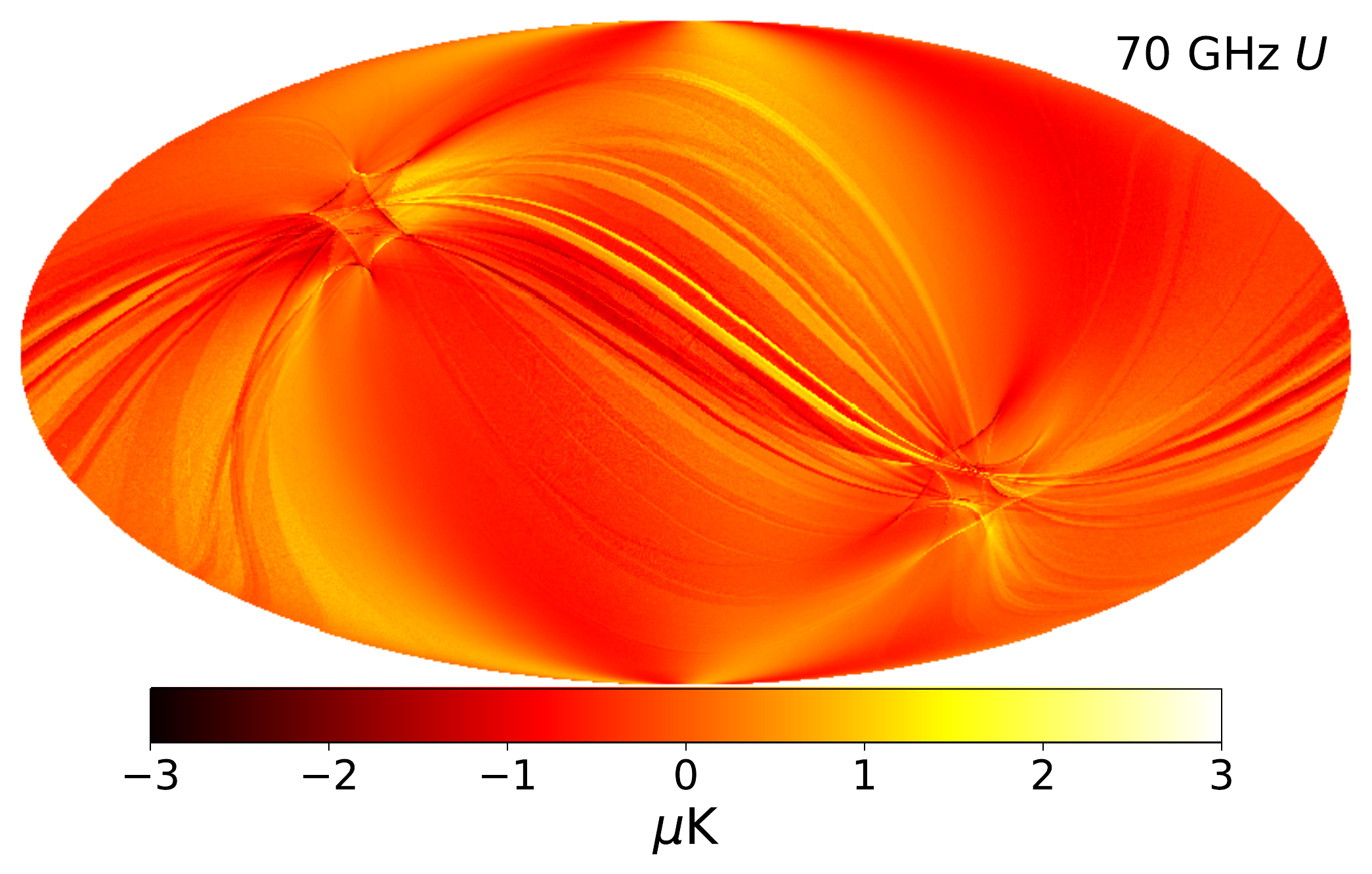}\\
  \caption{Maps of the sidelobes convolved with the sky at each of the three LFI frequencies. From top to bottom: 30\,GHz, 44\,GHz and 70\,GHz. The left column is the unpolarized sky signal, the central column is the Q polarization and the right column is U. Note the difference in the colour scales required to see the same level of detail in all three channels. 
  }\label{fig:slmean}
\end{figure*}

The treatment of the sidelobes is also important while generating orbital dipole and quadrupole estimates. Because \Planck\ is calibrated primarily from the dipole measurements \citep{planck2016-l01,npipe,BP07}, the sidelobe's contribution to the dipole can directly result in an absolute calibration error if not handled appropriately. While the CMB Solar dipole can easily be handled using the \texttt{Conviqt} approach described in Sec.~\ref{sec:conviqt}, the orbital dipole is not sky-stationary and thus must be handled separately. 

\BP\ generates orbital dipole and quadrupole estimates directly from the \Planck\ pointing information, using the satellite velocity data which has been stored at low resolution (one measurement per pointing period). With this information, it is possible to estimate the orbital dipole and quadrupole amplitude for each timestep, allowing the time-domain removal of the signal before it contaminates the final products with non-sky-stationary signal artifacts. Additionally, once this signal has been isolated from the raw data, it can be used as an aid in the calibration routines because of its highly predictable structure.

\BP\ uses the same technique as \Planck\ DR4 \citep[see appendix C]{npipe} to generate the orbital dipole and quadrupole estimate. That is, we express the signal $\tilde{D}$ seen by a detector observing a fixed direction $\hat{\boldsymbol n}_0$ as the convolution of the dipole and quadrupole signal on the sky, $D(\hat{\boldsymbol n})$, with the full $4\pi$ beam response, $B(\hat{\boldsymbol n}, \hat{\boldsymbol n}_0)$,
\begin{equation}
\tilde{D}(\hat{\boldsymbol n}_0) = \int d\Omega\, B(\hat{\boldsymbol n}, \hat{\boldsymbol n}_0) D(\hat{\boldsymbol n}).
\end{equation}
Here it is useful to break the dipole signal up into three orthogonal components in the standard Cartesian coordinates $(x,y,z)$, and we adopt the convention that the main beam points toward the north pole in our coordinate system.

The orbital CMB dipole and quadrupole can be expressed as a Doppler shift in each direction \citep{Notari:2015},
\begin{equation}
D(\hat{\boldsymbol n}) = T_0\left[ \beta \cdot \hat{\boldsymbol n}(1 + q\boldsymbol  \beta \cdot \hat{\boldsymbol n}) \right],
\label{eq:dipole_simple}
\end{equation}
where $\beta$ is the satellite velocity divided by the speed of light $\beta = \frac{\mathrm{v}}{c}$, $T_0$ is the CMB temperature and $q$ is quadrupole factor dependent on the frequency $\nu$, defined by 
\begin{equation}
q = \frac{a}{2} \frac{e^a + 1}{e^a -1}, \ \ \  \textrm{where} \ \ \ a = \frac{h\nu}{k_B T_0}.
\end{equation}
Inserting these expressions into Eq.~(\ref{eq:dipole_simple}), one obtains
\begin{equation}
\begin{split}
\tilde{D} = T_0 \int d\Omega_{\hat{\boldsymbol n}} B(\hat{\boldsymbol n}, \hat{\boldsymbol n}_0) \left[ x \ \beta_x + y\ \beta_y +  z\ \beta_z +\right. \\
q\left(x^2\ \beta^2_x + y^2\ \beta^2_y + z^2\ \beta^2_z + \right. \\
\left. \left. 2xy\ \beta_x\beta_y + 2xz\ \beta_x\beta_z + 2yz\ \beta_y\beta_z\right) \right],\label{eq:dipole}
\end{split}
\end{equation}
where $\hat{\boldsymbol n} = (x,y,z)$ is a unit direction vector that is also the integration variable, $\hat{\boldsymbol n}_0$ is the fixed direction of the satellite pointing for this timestep. Noting that the geometric factors in this expression may be precomputed as
\begin{equation}
S_{x} = \int x\, B(\hat{\boldsymbol n}, \hat{\boldsymbol n}_0)\, d\Omega_{\hat{\boldsymbol n}},
\end{equation}
we see that Eq.~\eqref{eq:dipole} may be written in the following form,
\begin{align}
\tilde{D} &= T_0 \left[ S_x \ \beta_x + S_y\ \beta_y +  S_z\ \beta_z +\right. \nonumber\\
  &\quad\quad\,\,\,\, q\left(S_{xx}\ \beta^2_x + S_{yy}\ \beta^2_y + S_{zz}\ \beta^2_z + \right. \nonumber\\
  &\quad\quad\quad\,\,\left.\left.2S_{xy}\ \beta_x\beta_y + 2S_{xz}\ \beta_x\beta_z + 2S_{yz}\ \beta_y\beta_z\right) \right]\label{eq:dipole_fast}.
\end{align}
To compute $\tilde{D}$ for an arbitrary beam orientation, one simply needs to rotate the satellite pointing and velocity vectors into the coordinate system used to define $S$, and then one can evaluate Eq.~\eqref{eq:dipole_fast} very quickly. 

\begin{figure*}[t]
  \center
  \includegraphics[width=0.33\linewidth]{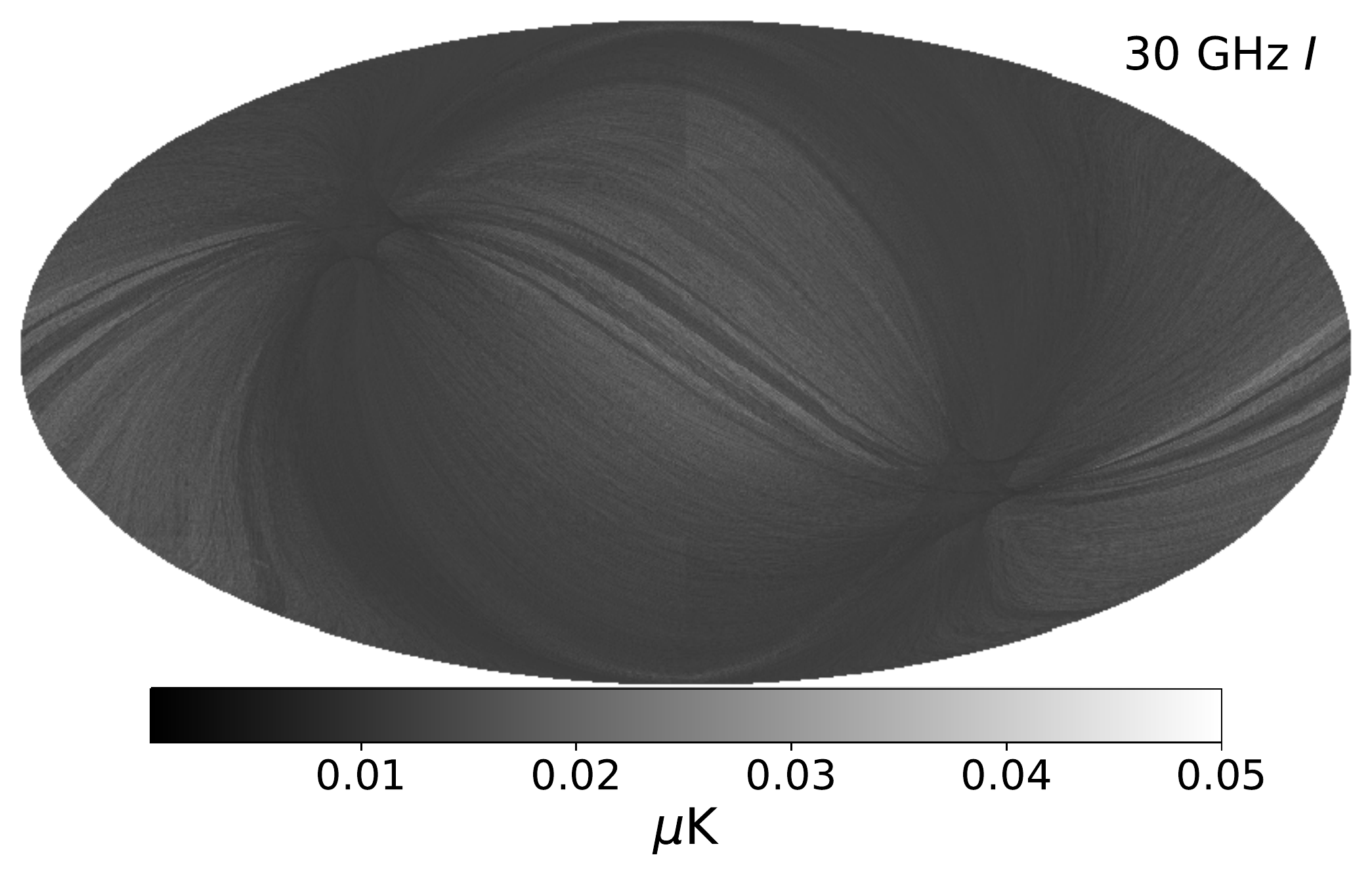}
  \includegraphics[width=0.33\linewidth]{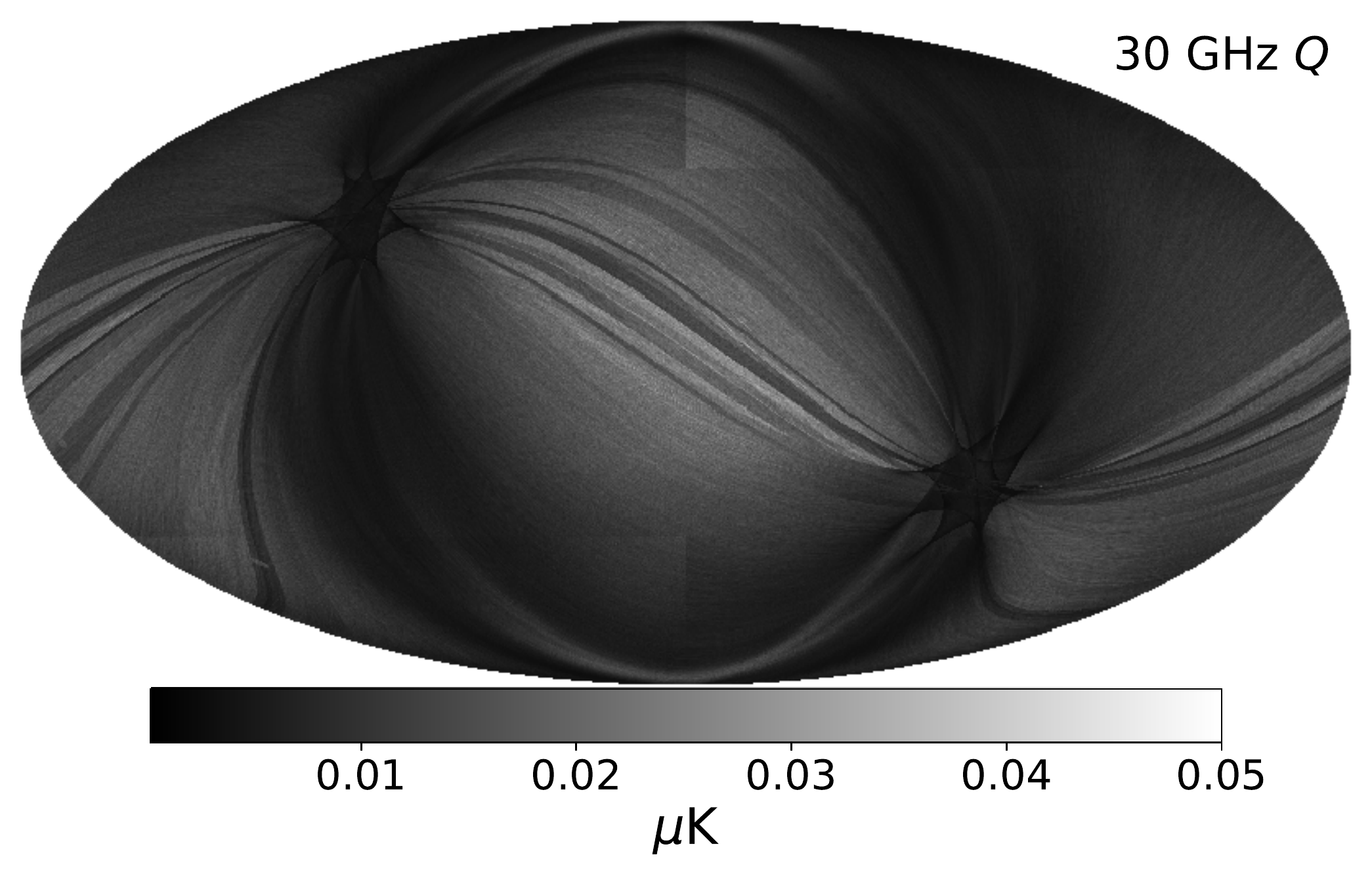}
  \includegraphics[width=0.33\linewidth]{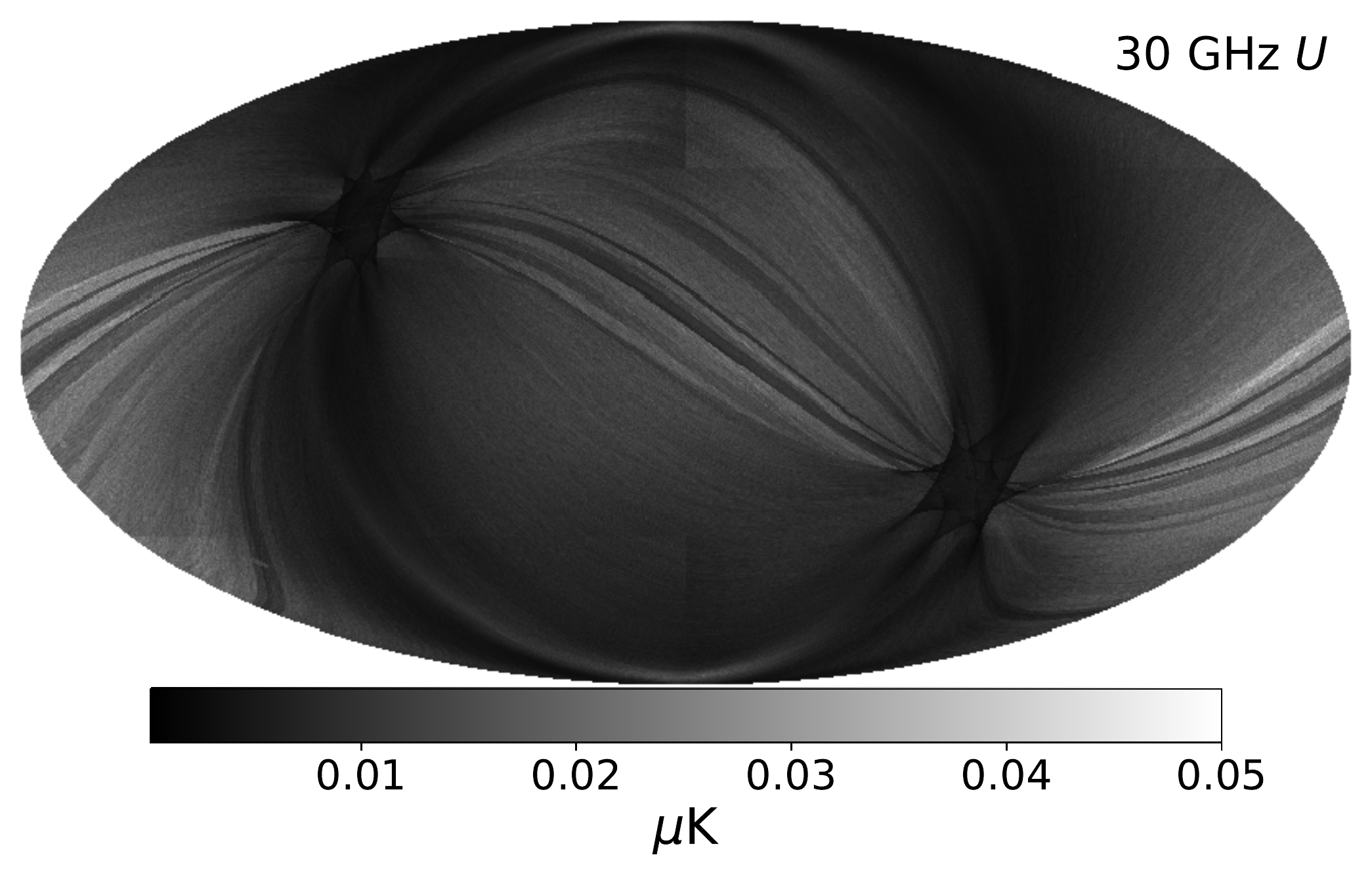}\\
    \includegraphics[width=0.33\linewidth]{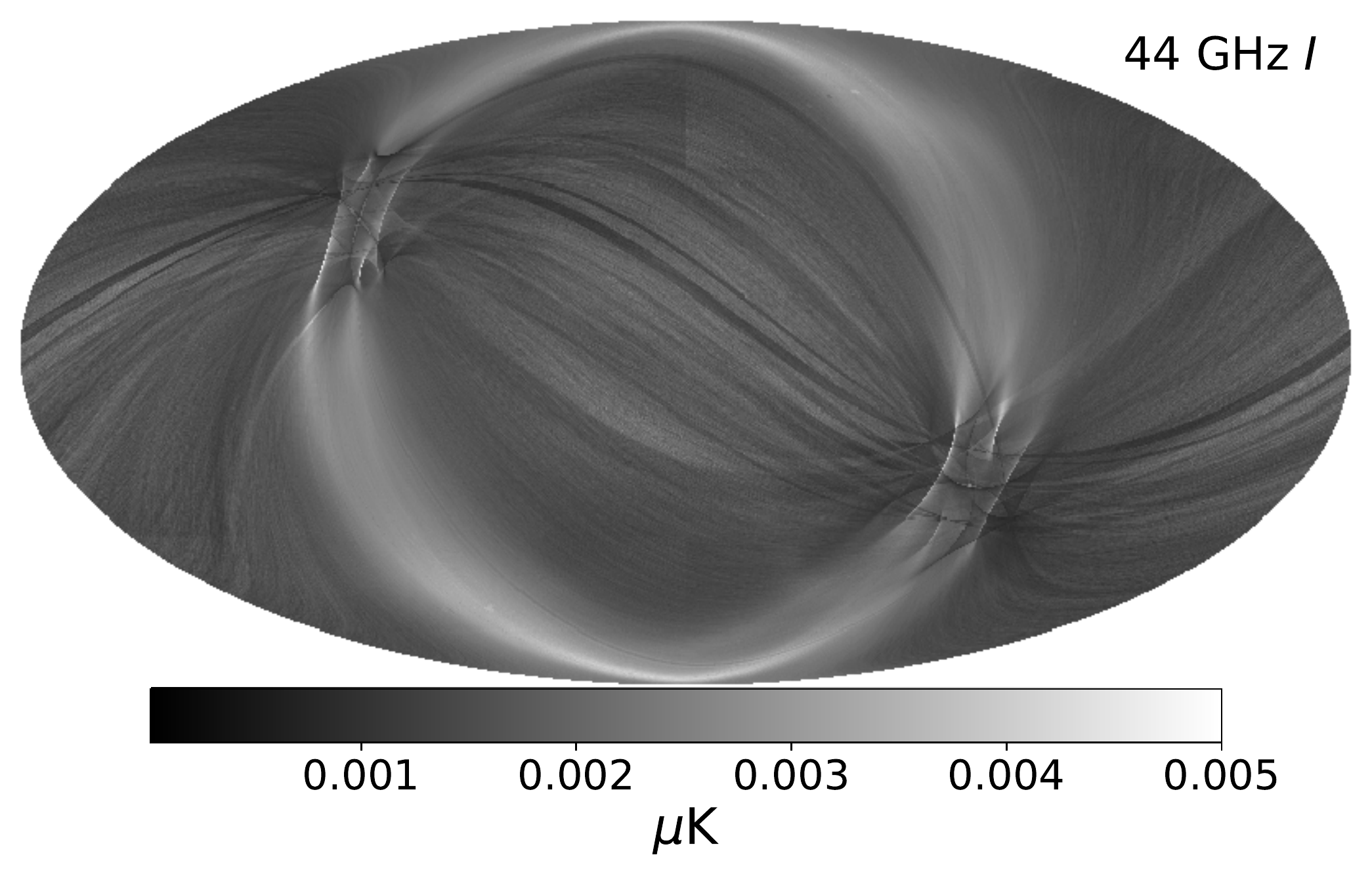}
  \includegraphics[width=0.33\linewidth]{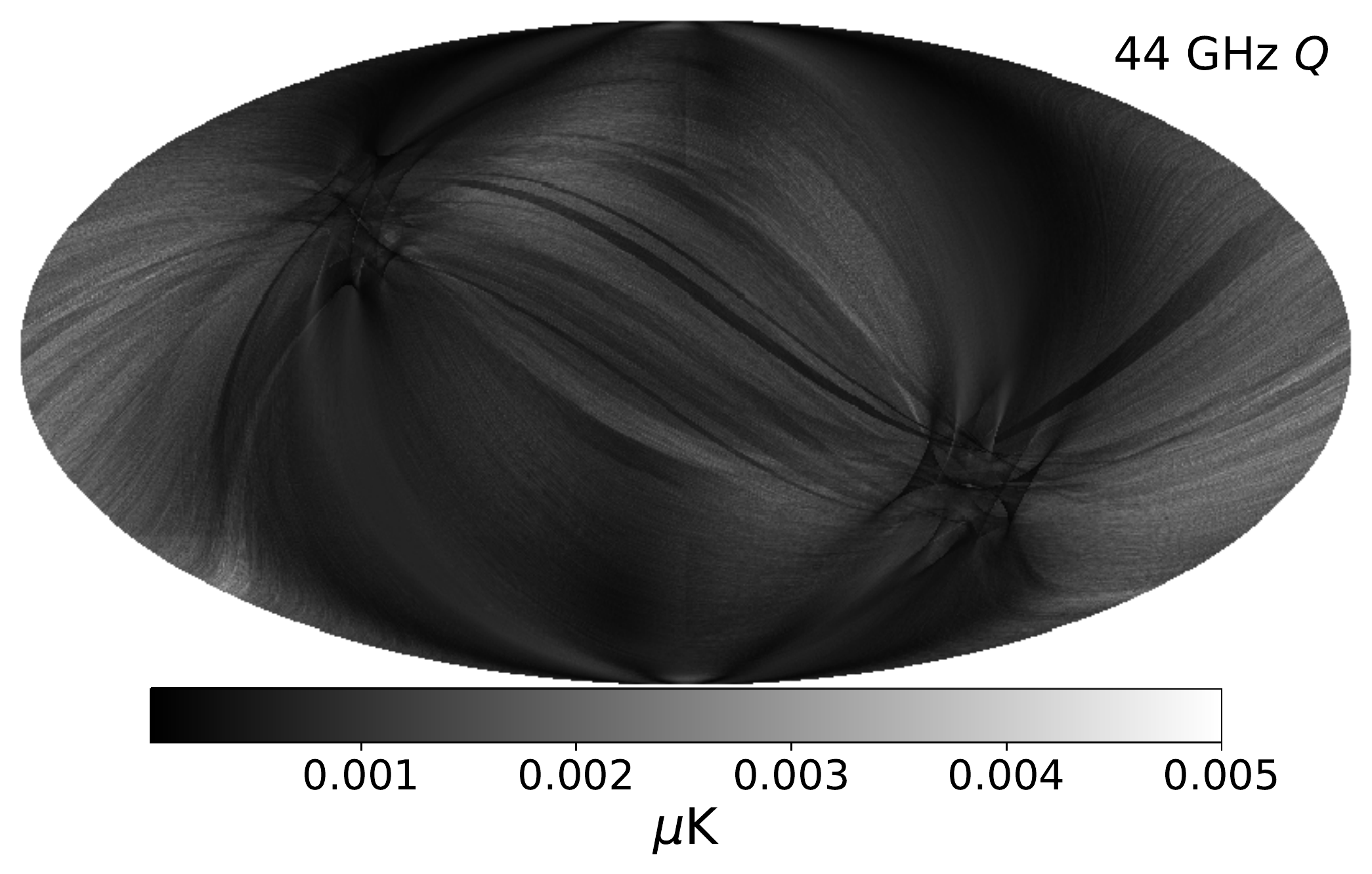}
  \includegraphics[width=0.33\linewidth]{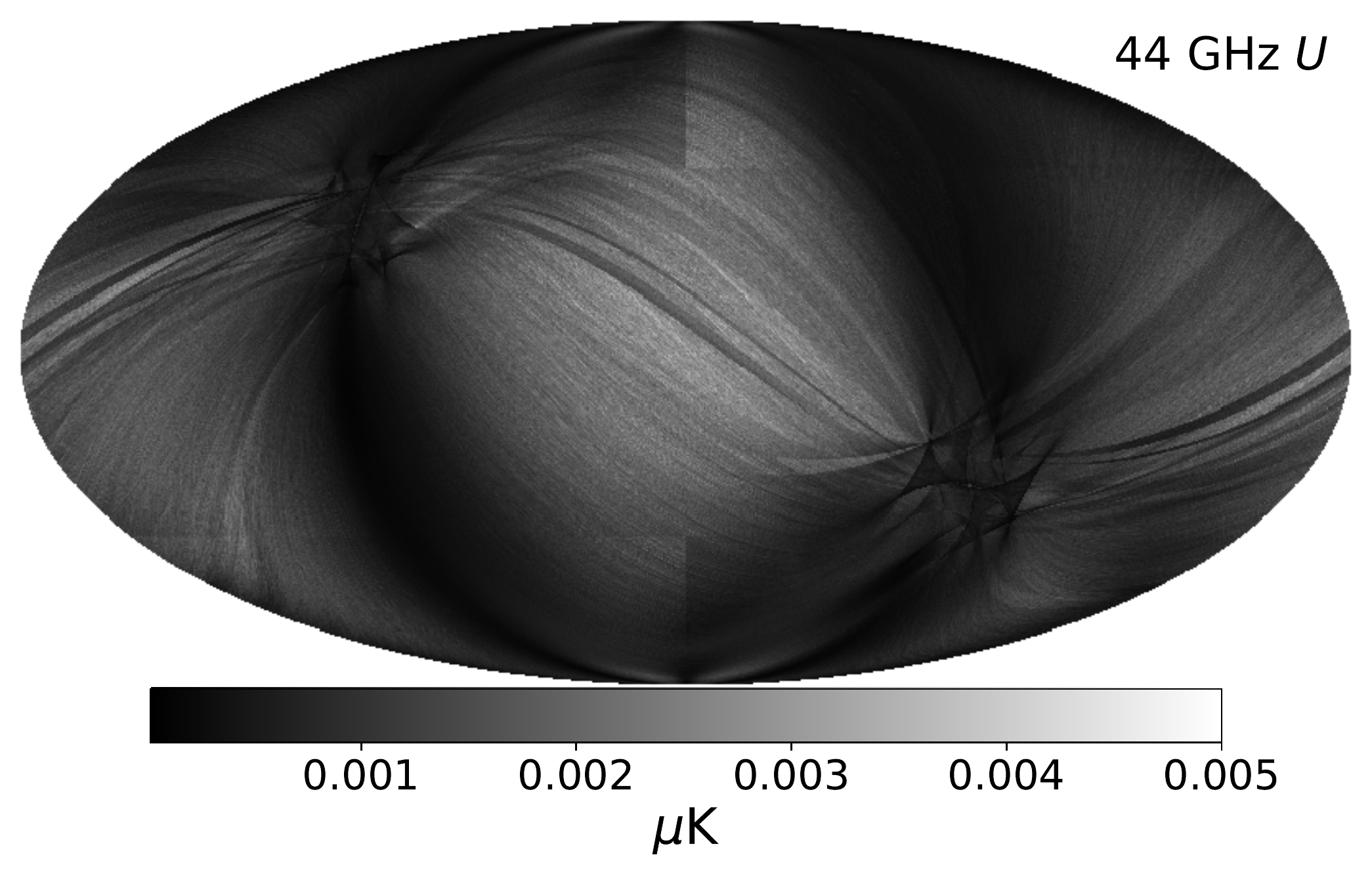}\\
    \includegraphics[width=0.33\linewidth]{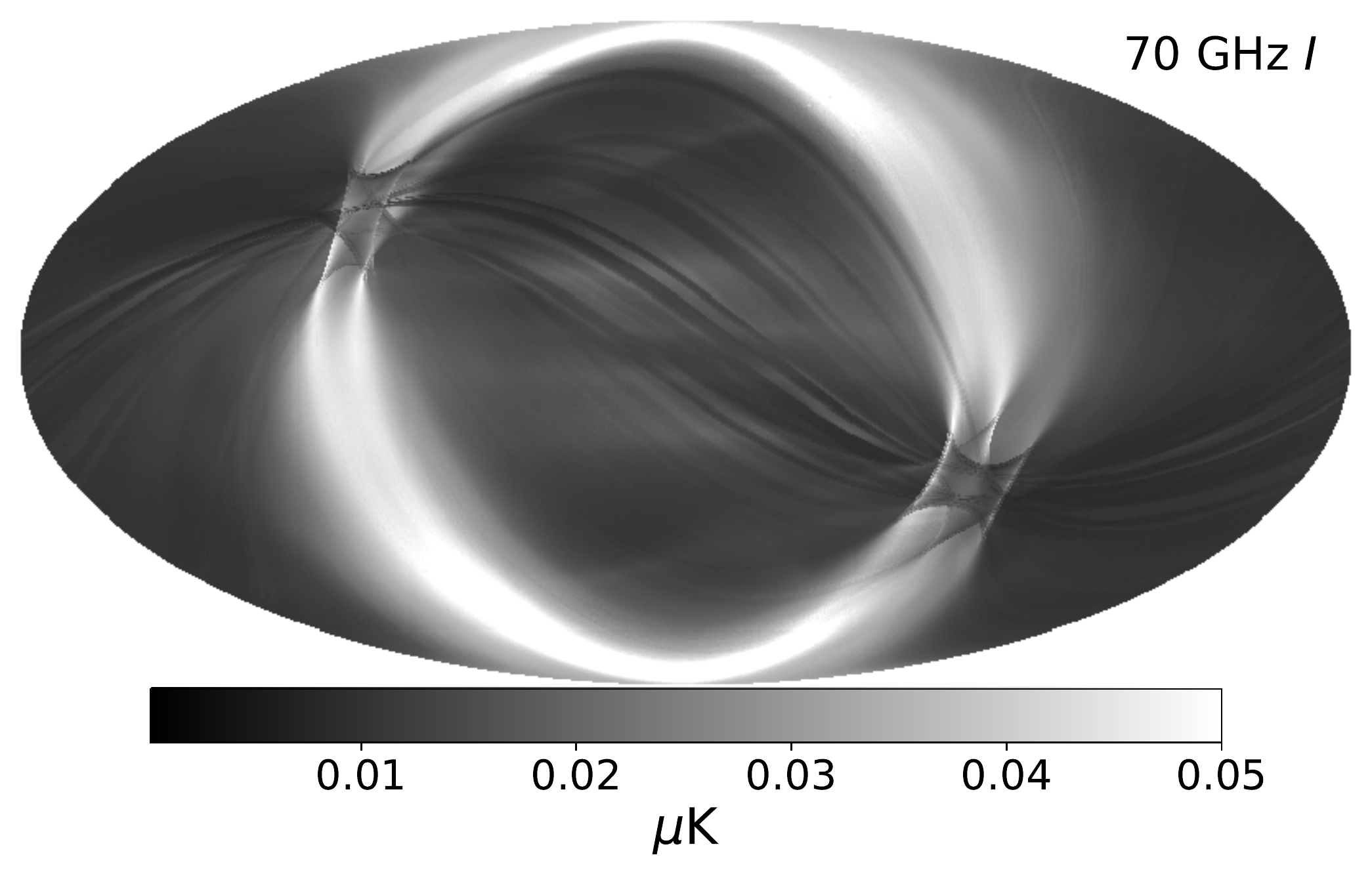}
  \includegraphics[width=0.33\linewidth]{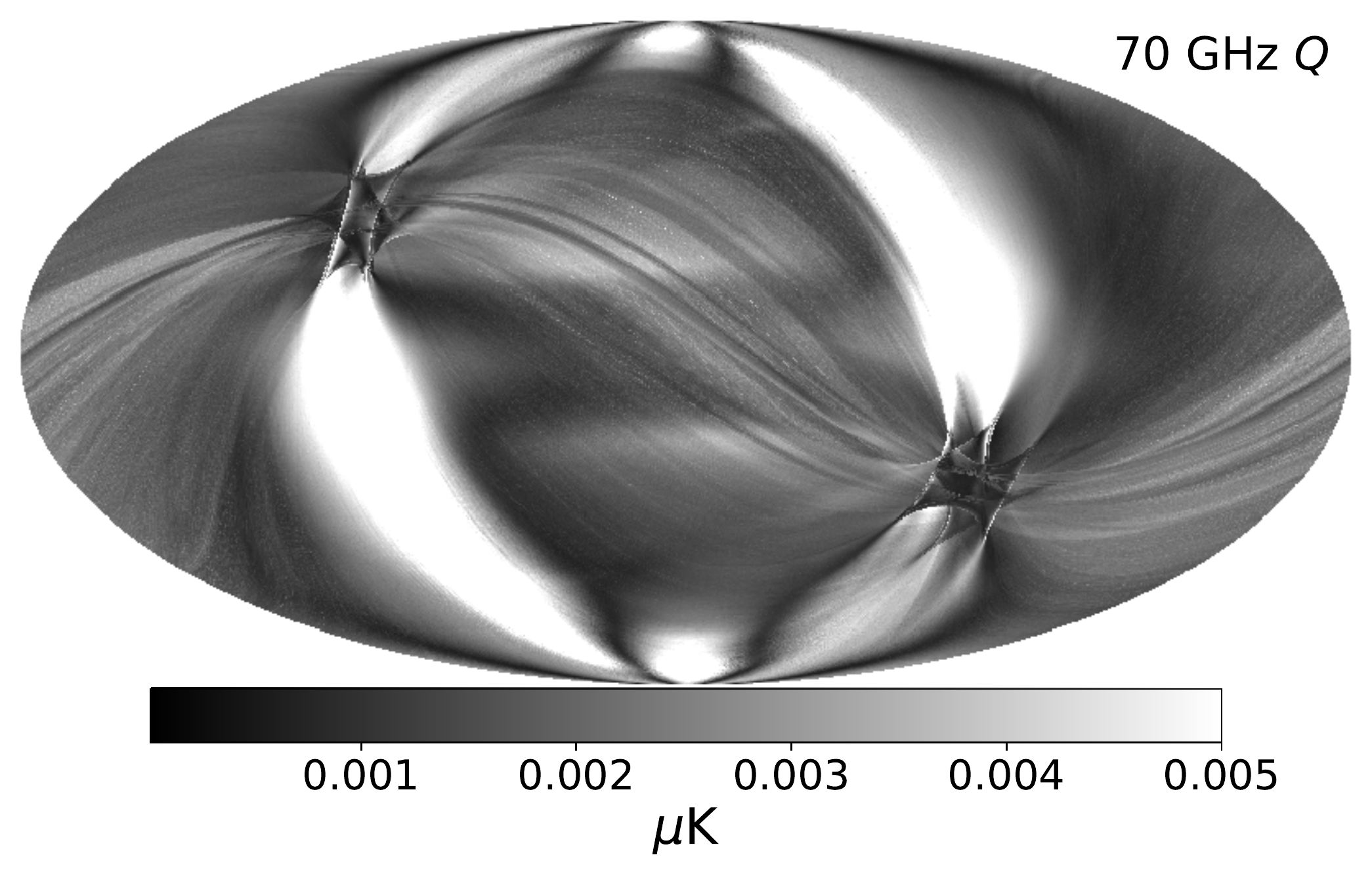}
  \includegraphics[width=0.33\linewidth]{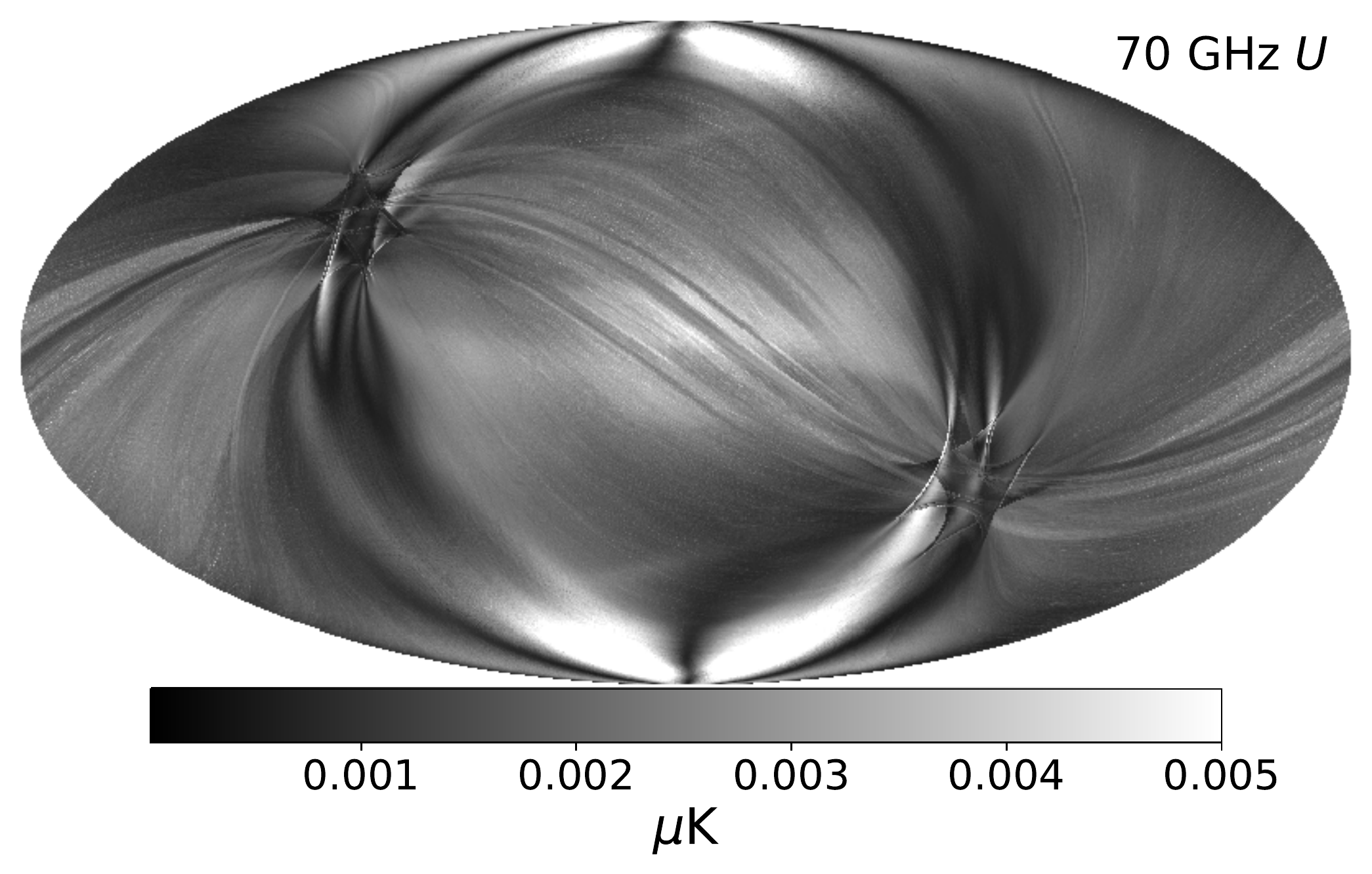}\\
  \caption{Sidelobe rms maps at each of the three LFI frequencies. From top to bottom: 30\,GHz, 44\,GHz and 70\,GHz. The left column is the unpolarized sky signal, the central column is the Q polarization and the right column is U. Again note the different colour scales.
  }\label{fig:slrms}
\end{figure*}

\BP\ further accelerates this operation by computing this rotation for only one point in twenty (chosen so as to still fully sample the dipole), and using a spline to interpolate between them. This saves the cost of calculating a new rotation matrix at each step, and instead relies on the smoothness of the signal to ensure continuity. The algorithm treats the final few points of each pointing period that do not divide evenly into the subsampling factor separately. This allows the use of regular bin widths, which greatly speeds up the splining routines, while the final few points are calculated using the slower rotation matrix technique. 

\begin{figure}[t]
  \center
  \includegraphics[width=0.95\linewidth]{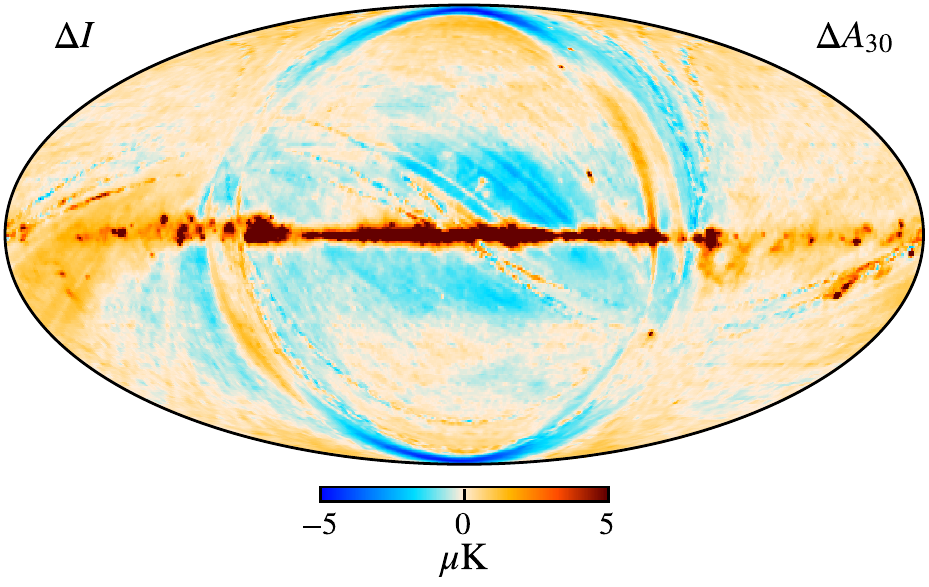}\\
  \includegraphics[width=0.95\linewidth]{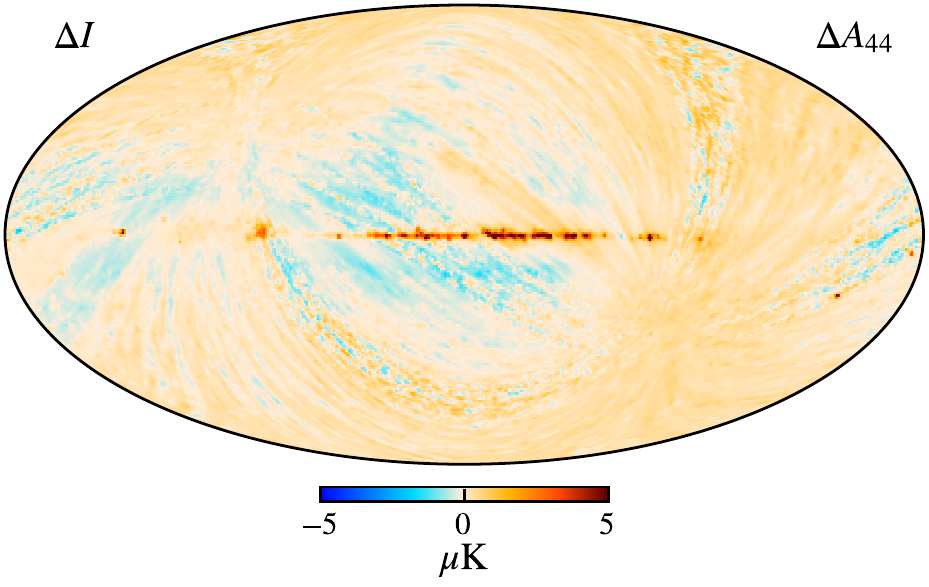}\\
  \includegraphics[width=0.95\linewidth]{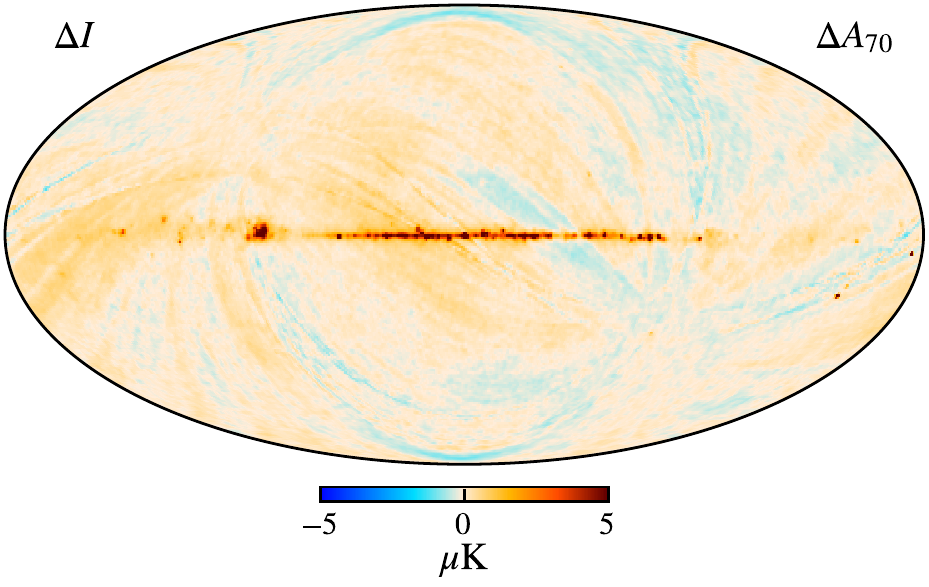}
  \caption{Frequency map difference plots smoothed to one degree at (top to bottom) 30, 44 and 70\,GHz, comparing two pipeline executions with the same seed, one of which has no sidelobe correction.
  }\label{fig:freqdiff}
\end{figure}

\begin{figure}[t]
  \center
  \includegraphics[width=0.95\linewidth]{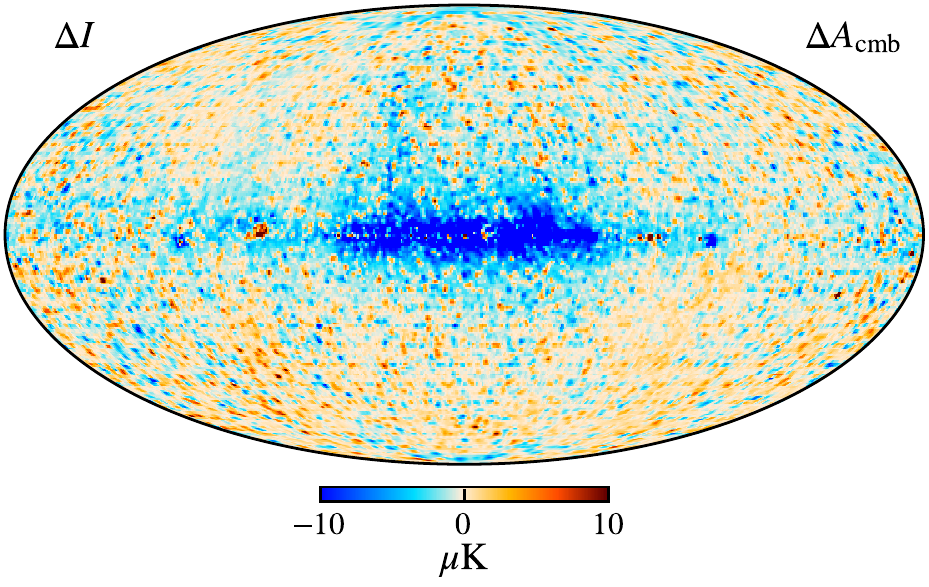}\\
  \includegraphics[width=0.95\linewidth]{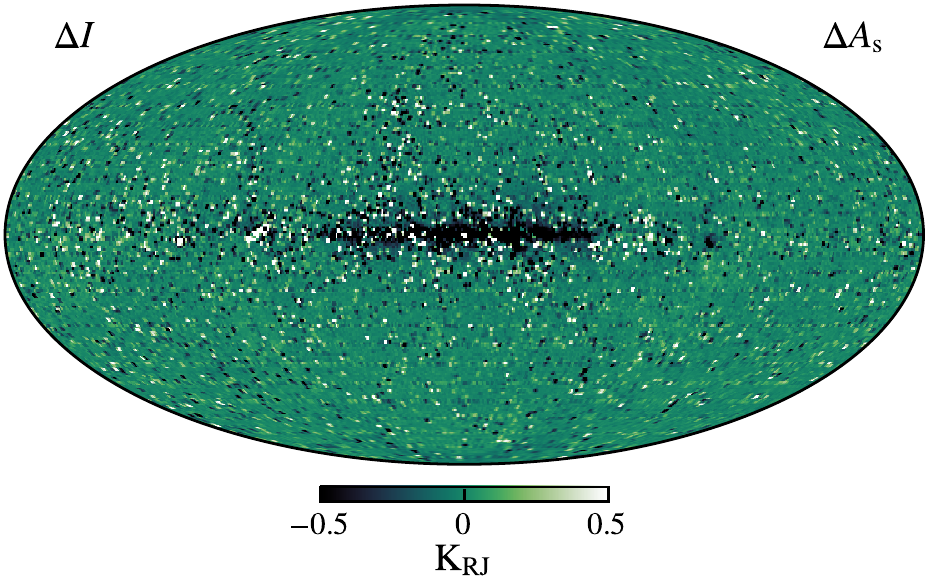}\\
  \includegraphics[width=0.95\linewidth]{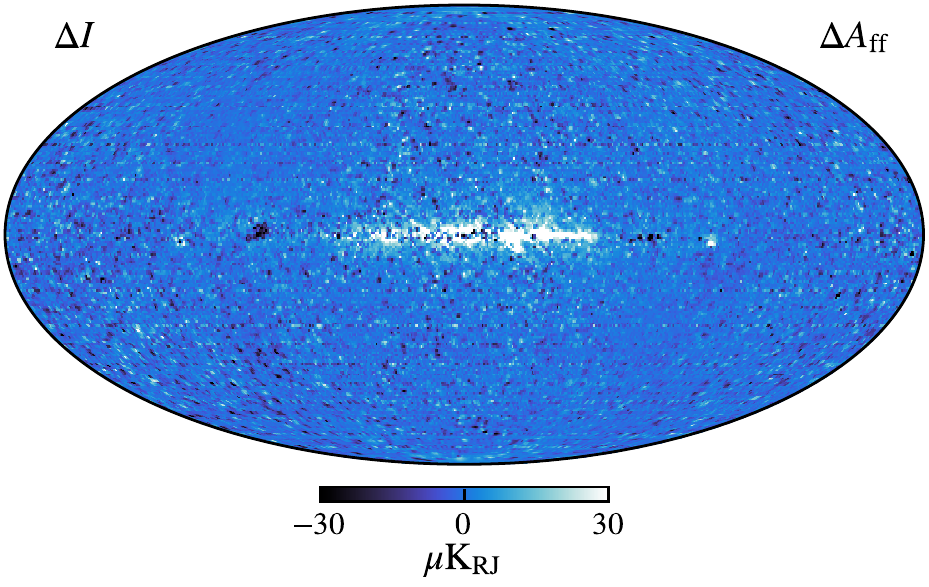}\\
  \includegraphics[width=0.95\linewidth]{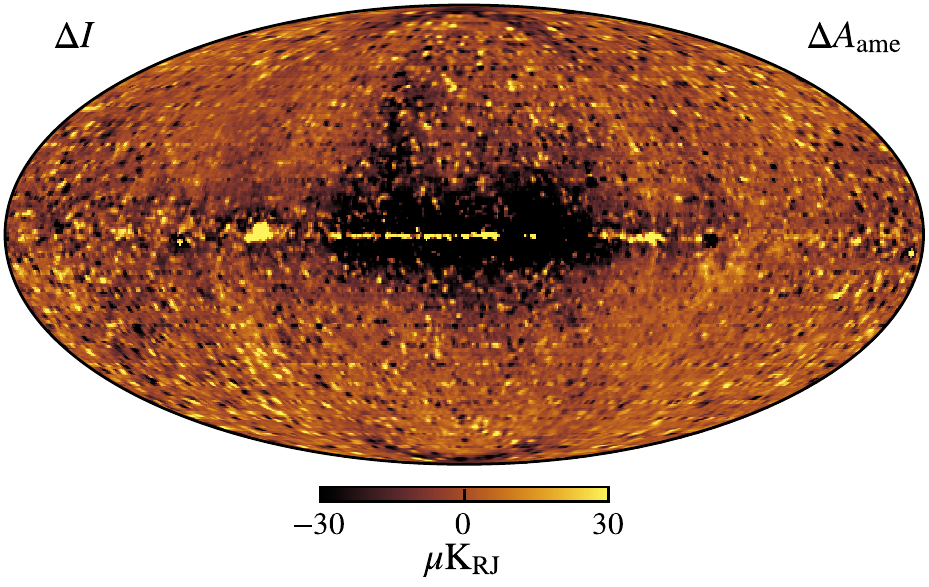}
  \caption{Component map difference plots for (top to bottom) CMB, synchrotron, AME and freefree emission, comparing two pipeline executions with the same seed, one of which has no sidelobe correction.
  }\label{fig:compdiff}
\end{figure}

\section{Sidelobe Estimates}
\subsection{Posterior mean corrections}

Figure~\ref{fig:slmean} shows the mean sidelobe signal estimates at each of the three LFI frequencies for the entire mission, co-added across each frequency and projected into sky coordinates, identically to the way the true sky signal is treated. Each map is averaged over 90 Gibbs samples produced in the main \BP\ analysis \citep{BP01}, after discarding burn-in and thinning the remaining chain by a factor of ten.

We note that these maps follow the traditional \Planck\ LFI method of sidelobe correction by producing these signals in the time domain during TOD processing. These templates are therefore exactly correct for the maps produced by these pipeline runs, but will not match precisely with analyses that use different data cuts, flagging or channel selection.

These results look visually similar to the corresponding \Planck\ DPC results presented in Fig.~7 of \citet{planck2014-a04}. The main difference is that the current results also include the CMB dipole, whereas the LFI 2015 DPC analysis showed the sidelobe pickup of dipole-subtracted maps. We see that the sidelobe signal is strongest at 30\,GHz, and that the dominant features in the co-added sky maps consist of a series of rings created by the interplay between the sidelobe pickup and bright Galactic plane features. 

Figure~\ref{fig:slmean} also clearly indicates that the overall level of sidelobe pickup at 44\,GHz is significantly lower than for the 30 and 70\,GHz channels. This is due to the particular location in the focal plane of two of the three 44\,GHz feedhorns, which results in a significant under-illumination of both the primary and secondary reflectors of the Planck telescope for those two horns (see Fig.~4 of \cite{sandri2010}).

\subsection{Error Propagation}

In addition to the posterior mean sidelobe maps, the \BP\ pipeline outputs also provide an estimate of the sidelobe stability and statistical variation. Figure~\ref{fig:slrms} shows the rms maps generated from the same sample of sidelobe signal estimates as was used in Fig.~\ref{fig:slmean}. Clear evidence of the scanning pattern can be seen, which is expected. The sharp vertical lines visible in polarization (clearest in 30\,GHz Q and U at the top, and 44\,GHz $U$ at the top and bottom) have been previously examined by the \Planck\ team, and are caused by a chance alignment between the non-dense \Planck\ scanning strategy and the shape of the \healpix\ pixels. For an example of this effect, see Fig.~15 of \citet{planck2013-p03c}.

These posterior rms maps cannot be considered true sidelobe error estimates, however, as they do not account for uncertainties in the sidelobe response itself. Rather, they only show the change in the estimated sidelobe signal due to sky model variations from component separation. Full sidelobe error propagation would require sampling over the physical parameters that determine the detectors' sidelobe response on the sky. While sampling the full set of optical model parameters is likely to be infeasible due to excessive computational time, identifying a minimal parameter set that may account for the main potential variations in the sidelobe response functions, and precomputing response functions over a grid of such parameters, would result in physically motivated uncertainties for the sidelobe models. This approach will be developed for future applications such as the \LiteBIRD\ mission \citep{LiteBIRD}.

\section{Impact on Frequency and Component Maps}

To assess the importance of sidelobe corrections on frequency and component maps, we perform two runs of the  \commander\ code, starting from the same input data as the main pipeline run, and with identical random seeds. As a comparison, we remove the far sidelobe correction from one of these secondary pipeline executions, and we difference the results between these two pipelines. 

Figure~\ref{fig:freqdiff} shows the differences in the frequency maps between the two cases in temperature, where the effects are the most obvious. The only large-scale features that can clearly be seen are the small dipole differences (most clearly visible at 70\,GHz). These are directly caused by the dipolar component seen in Fig.~\ref{fig:slmean}, as this contribution to the total sky signal that was in the sidelobe term is now unaccounted for. In previous analyses, these dipole contributions were handled through specific modeling of exactly these effects, but this test makes it explicitly clear that correct dipole measurements require accurate knowledge of the sidelobe pickup.

Next, we see two more features in the difference maps that are more localized. The first of these are the ring structures that match the actual sidelobe map structures quite closely. These are of course the same rings from Fig.~\ref{fig:slmean}, which are not accounted for in the second pipeline run without sidelobe corrections. Additionally, there are some uniform residuals that are visible in the Galactic plane regions of the difference maps. These are caused by calibration mismatch between the detectors at a single frequency. As each of the detectors now sees a slightly different dipole signal on the sky, depending on its specific sidelobe response, their calibrations do not agree with one another, which causes signal residuals which are most visible in the plane where the signal amplitude is highest.

Figure~\ref{fig:compdiff} shows the differences in component maps from this same comparison, again in temperature. The CMB as well as the three low-frequency foreground components are estimated using the standard \commanderthree\ technique described in \cite{bp13}. The AME component sees similar issues to the ones seen by the frequency maps above. The dipole is slightly incorrect, there are sidelobe-esque stripes, and the Galactic plane shows a strong residual, all of which are effects that have been seen directly in the frequency maps. The dipole difference seen here is precisely the one that contributes to the difference in calibration between the two different pipeline executions. 

The other three components (synchrotron, CMB and free-free) show relatively less structural difference. They have absorbed some of the sidelobe-like ring structures, but the primary difference can be seen most clearly in the Galactic plane. Here, we notice a large residual caused by the inaccurate model of the Galactic emission being altered slightly by the gain and calibration differences between the two runs. As the Galactic emission is significantly brighter than the rest of the sky, small changes in calibration produce large errors like the ones seen here.

\begin{figure}[t]
  \center
  \includegraphics[width=0.95\linewidth]{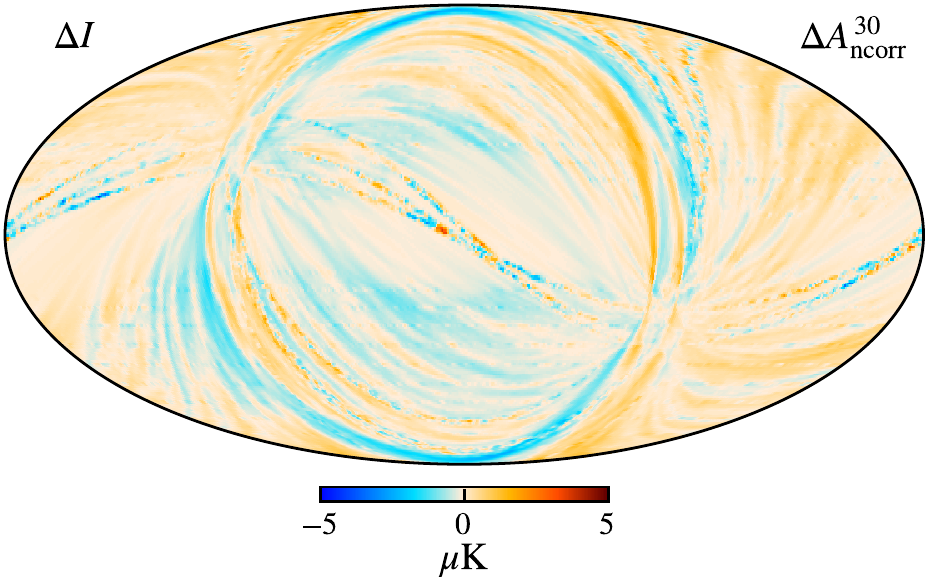}\\
  \caption{Difference in correlated noise, projected into the map domain at 30\,GHz comparing two pipeline executions with the same seed, one of which has no sidelobe correction.
  }\label{fig:ncorr}
\end{figure}

Finally, in Fig.~\ref{fig:ncorr}, we show the correlated noise map difference at 30\,GHz between the two runs. Here, we see that some of the missing sidelobe signal has been accommodated by the correlated noise component. These structures mirror the strongest sidelobe-like signals in the 30\,GHz difference map in the top panel of Fig.~\ref{fig:freqdiff}. The fact that $n_{\mathrm{corr}}$ can accommodate some sidelobe residuals is in fact helpful, as it allows for some leeway in the final sidelobe model, in the sense that small uncertainties and artifacts that are inconsistent between different frequency channels will be mostly absorbed into the correlated noise component, rather than in the sky maps. The differences that we see in the maps, however, indicate that this process is not perfect, as some of the spurious signal still makes it to the final maps without a perfect sidelobe model. For a real-world example of these issues, we refer the interested reader to the on-going \BP\ re-analysis of the \WMAP\ data, for which far sidelobe contamination appears to be a dominant problem \citep{bp17}.

The residual errors seen in Figs.~\ref{fig:freqdiff} and \ref{fig:compdiff} are also present in the \BP\ analysis, albeit at much lower levels. We know that our knowledge and modeling of the sidelobes are imperfect, as they are based on limited measurements of the physical LFI sidelobes, and some of the power is unaccounted for. Future applications of the pipeline that aim for a robust $r\le 0.01$ measurement will be required to marginalize over the sidelobe uncertainties in some manner, either by directly Gibbs sampling a subset of the instrument parameters or by parameterizing and fitting sidelobe error estimates. We do not believe that the sidelobe contribution causes significant errors in the LFI sample sets produced by \BP, as it is unlikely to be more than a 10\,\% error on the sidelobe estimates of Fig. \ref{fig:slmean}. At 30\,GHz, this corresponds to at most a 0.05\,\% error in our temperature maps and a 1\,\% error in polarization. We do expect however, that as instrumental sensitivities improve, especially in polarization, this sidelobe term will need to be modeled very accurately, and the corresponding uncertainties must be propagated properly, for instance using methods similar to those presented in this paper.

\section{Summary and Conclusions}
\label{sec:conclusions}

This paper presents a formulation of the \texttt{Conviqt} algorithm in terms of spin-weighted spherical harmonics. This algorithm is already implemented in the latest versions of the \texttt{libconviqt} library, and it has now also been re-implemented directly into \commander, where it is used for sidelobe corrections for the \BP\ analysis framework. Based on the Monte Carlo samples produced in that analysis, we have presented novel posterior mean and standard deviation maps for each of the three \Planck\ LFI frequency bands.

The full-sky sidelobe treatment techniques presented here are easily generalizable to other experiments, and can be tuned to match the required spatial characteristics of other instruments simply by adjusting the spherical harmonic bandpass parameters, $l_\mathrm{max}$ and $m_\mathrm{max}$, of the sidelobe description. The only requirement for using the code with a new instrument is a \healpix-compatible description of the sidelobe response function per detector. The more accurate this characterization of the instrument is, the better the sidelobe estimate will approximate the true sidelobe contamination in the timestream.

We note that the approach presented here is less useful for ground or balloon based experiments where the dominant sidelobe pickup contains radiation from an environmental source. This pickup is not sky-synchronous, and thus cannot be modeled purely as a beam-sky convolution, but must include additional contributions from, for example, telescope baffles, ground pickup or clouds. For these types of experiments, other techniques such as aggressive baffling are likely better suited.

We also stress that the current implementation only supports sidelobe error propagation for sky model uncertainties, not uncertainties in the actual sidelobe response function itself, and these are very likely to dominate the total sidelobe error budget. Future work should therefore aim to introduce parametric models for the sidelobe response itself, and sample (or, at least, marginalize) over the corresponding free parameters, as these are typically one of the most important unknowns for many experiments.

Finally, we note that future CMB experiments such as \LiteBIRD, which are targeting low $B$-mode limits, may need to consider more complex ways of handling sidelobes and beams. The ultimate solution in this respect is $4\pi$ beam convolution for every single timestep, which could be achieved using a similar framework to the approach discussed here. This would remove the sidelobes as a nuisance signal from the data model of Eq.~(\ref{eq:datamodel}) and instead incorporate them directly into the beam term, $\B_{pp',j}$. This approach should be feasible for a relatively low-resolution experiment such as \LiteBIRD, and will be investigated going forward.

\begin{acknowledgements}
  We thank Prof.\ Pedro Ferreira and Dr.\ Charles Lawrence for useful suggestions, comments and 
  discussions. We also thank the entire \Planck\ and \WMAP\ teams for
  invaluable support and discussions, and for their dedicated efforts
  through several decades without which this work would not be
  possible. The current work has received funding from the European
  Union’s Horizon 2020 research and innovation programme under grant
  agreement numbers 776282 (COMPET-4; \BP), 772253 (ERC;
  \textsc{bits2cosmology}), and 819478 (ERC; \textsc{Cosmoglobe}). In
  addition, the collaboration acknowledges support from ESA; ASI and
  INAF (Italy); NASA and DoE (USA); Tekes, Academy of Finland (grant
   no.\ 295113), CSC, and Magnus Ehrnrooth foundation (Finland); RCN
  (Norway; grant nos.\ 263011, 274990); and PRACE (EU).
\end{acknowledgements}

\bibliographystyle{aa}

\bibliography{Planck_bib,BP_bibliography,sources}

\end{document}